\documentclass[12pt,english,reprint,amsmaths,amssymb,prl]{revtex4-1}
\usepackage[T1]{fontenc}
\usepackage[latin9]{inputenc}
\usepackage{color}
\usepackage{babel}
\usepackage{array}
\usepackage{refstyle}
\usepackage{booktabs}
\usepackage{multirow}
\usepackage{amstext}
\usepackage{amssymb}
\usepackage{graphicx}
\usepackage[unicode=true,pdfusetitle,
 bookmarks=false,
 breaklinks=false,pdfborder={0 0 0},pdfborderstyle={},backref=false,colorlinks=true]
 {hyperref}

\makeatletter

\AtBeginDocument{\providecommand\subsecref[1]{\ref{subsec:#1}}}
\AtBeginDocument{\providecommand\figref[1]{\ref{fig:#1}}}
\AtBeginDocument{\providecommand\tabref[1]{\ref{tab:#1}}}
\providecommand{\tabularnewline}{\\}
\newcommand{\lyxdot}{.}

\RS@ifundefined{subsecref}
  {\newref{subsec}{name = \RSsectxt}}
  {}
\RS@ifundefined{thmref}
  {\def\RSthmtxt{theorem~}\newref{thm}{name = \RSthmtxt}}
  {}
\RS@ifundefined{lemref}
  {\def\RSlemtxt{lemma~}\newref{lem}{name = \RSlemtxt}}
  {}

\renewcommand{\tabref}{\Tabref}
\renewcommand{\figref}{\Figref}

\usepackage[shortcuts]{extdash}

\makeatother

\begin{document}

\title{Ferroelectric ZrO$_{2}$ monolayers as buffer layers between SrTiO$_{3}$
and Si}

\author{Mehmet Dogan{*}$^{1,2,3,4}$ and Sohrab Ismail-Beigi$^{1,2,5,6}$}

\affiliation{$^{1}$Center for Research on Interface Structures and Phenomena,
Yale University, New Haven, Connecticut 06520, USA $\linebreak$ $^{2}$Department
of Physics, Yale University, New Haven, Connecticut 06520, USA $\linebreak$
$^{3}$Department of Physics, University of California, Berkeley,
California 94720, USA $\linebreak$ $^{4}$Materials Science Division,
Lawrence Berkeley National Laboratory, Berkeley, California 94720,
USA $\linebreak$ $^{5}$Department of Applied Physics, Yale University,
New Haven, Connecticut 06520, USA $\linebreak$ $^{6}$Department
of Mechanical Engineering and Materials Science, Yale University,
New Haven, Connecticut 06520, USA $\linebreak$ {*}Corresponding author: mhmtdogan@gmail.com}

\date{March 27, 2019}
\begin{abstract}
A monolayer of ZrO$_{2}$ has recently been grown on the Si(001) surface
and shown to have ferroelectric properties, which signifies the realization
of the lowest possible thickness in ferroelectric oxides (M. Dogan
et al., \emph{Nano Lett., }\textbf{18 (1)} (2018) \citep{dogan2018singleatomic}).
In our previous computational study, we reported on the multiple (meta)stable
configurations of ZrO$_{2}$ monolayers on Si, and how switching between
a pair of differently polarized configurations may explain the observed
ferroelectric behavior of these films (M. Dogan and S. Ismail-Beigi,
\emph{arXiv}:1902.01022 (2019) \citep{dogan2019theoryof}). In the
current study, we conduct a DFT-based investigation of (i) the effect
of oxygen content on the ionic polarization of the oxide, and (ii)
the role of zirconia monolayers as buffer layers between silicon and
a thicker oxide film that is normally paraelectric on silicon, e.g.
SrTiO$_{3}$. We find that (i) total energy-vs-polarization behavior
of the monolayers, as well as interface chemistry, is highly dependent
on the oxygen content; and (ii) SrTiO$_{3}$/ZrO$_{2}$/Si stacks
exhibit multiple (meta)stable configurations and polarization profiles,
i.e. zirconia monolayers can induce ferroelectricity in oxides such
as SrTiO$_{3}$ when used as a buffer layer. This may enable a robust
non-volatile device architecture where the thickness of the gate oxide
(here strontium titanate) can be chosen according to the desired properties.
\end{abstract}
\maketitle

\section{Introduction\label{sec:Introduction}}

Metal oxide thin films exhibit a diverse set of physical phenomena
with technological applications, such as ferroelectricity, ferromagnetism
and superconductivity, and thus have motivated intense scientific
research for decades \citep{hwang2012emergent,mannhart2010oxideinterfacestextemdashan}.
One of these phenomena, thin film ferroelectricity, potentially enables
non-volatile devices such as ferroelectric field-effect transistors
(FEFET). In traditional field-effect transistors, the state of the
device is determined by the applied gate voltage, meaning when the
gate voltage is turned off, the state is also switched to ``off''
(hence it is volatile). In contrast, in a FEFET, the polarization
of the thin film oxide can be retained, and keeps the state ``on''
after the gate voltage is turned off (hence it is non-volatile). Encoding
the state in the oxide rather than the applied gate voltage greatly
decreases the energy requirement and increases the speed of these
transistors \citep{mckee2001physical,garrity2012growthand}. Achieving
this requires a metal oxide which remains (or becomes) ferroelectric
as a thin film on a semiconductor, and an interface between the two
materials that is atomically abrupt, causing the electronic states
between them to be coupled \citep{reiner2009atomically,reiner2010crystalline,dogan2017abinitio}.
The first of these prerequisites has been a primary challenge due
to the fact that bulk ferroelectrics do not retain their macroscopic
polarization under a critical thickness because of the depolarizing
field caused by surface bound charges \citep{batra1973phasetransition,dubourdieu2013switching}.
However, instead of focusing on bulk ferroelectrics, it is possible
to engineer atomically abrupt interfaces between semiconductors and
oxides such that the oxide is stable in multiple polarization states
in the thin film form \citep{dogan2017abinitio}. This approach utilizes
strong interface effects to offset the depolarizing effects of the
thin film's surface. Therefore, achieving an abrupt interface between
the oxide and the semiconductor is critical for both (i) coupling
their electronic states, and (ii) inducing multiple polarizations
in the oxide. This is challenging because of the amorphous oxide layers
(such as SiO$_{2}$) that usually form at the interfacial region \citep{robertson2006highdielectric,garrity2012growthand,mcdaniel2014achemical}.
However, recent developments in the growth methods such as molecular
beam epitaxy (MBE) allows us to overcome this difficulty in many materials
systems \citep{mckee1998crystalline,mckee2001physical,kumah2016engineered}.

In a recent letter, we reported on the ferroelectric behavior of atomically
thin ZrO$_{2}$ grown on Si(001) \citep{dogan2018singleatomic}. This
was achieved by atomic layer deposition (ALD) which produced an atomically
abrupt interface and a mostly amorphous oxide. Using amorphous Al$_{2}$O$_{3}$
as a top electrode, a gate stack was created, and ferroelectric behavior
was observed via $C-V$ measurements. In our following computational
work, we presented an in-depth analysis of the monocrystalline ZrO$_{2}$/Si(001)
interface \citep{dogan2019theoryof}. Using Monte Carlo simulations
on a discrete lattice model whose parameters are extracted from DFT
results, we conducted an investigation of the multi-domain film, which
approximates the experimental amorphous film. Our results suggested
that two low-energy configurations of opposite polarization may be
dominant in the experimental film. Thus the observed ferroelectric
behavior can be understood as locally switching between these configurations.

In this work, we present a complementary computational study of the
effect of oxygen content on the interface chemistry and the polarization
of these monolayers, and find that oxygen content can be used to adjust
the energy-vs-polarization behavior of these monolayers (\subsecref{ZrOx-on-Si}).
In addition, we investigate epitaxial SrTiO$_{3}$/ZrO$_{2}$/Si heterostructures
in order to test the idea that an ultrathin binary oxide such as ZrO$_{2}$
can induce ferroelectricity in a thicker perovskite oxide, such as
SrTiO$_{3}$, when used as a buffer layer (\subsecref{STO_ZrOx}).
We show that the SrTiO$_{3}$ thin films, paraelectric when grown
directly on Si(001) \citep{kolpak2011thermodynamic}, possess multiple
(meta)stable configurations with varying ionic polarization, in the
SrTiO$_{3}$/ZrO$_{2}$/Si stacks. These configurations are (meta)stable
for 1.5-3.5 u.c. of SrTiO$_{3}$. Therefore, it may be possible to
utilize these stacks in a non-volatile device such as a FEFET where
the thickness of the gate oxide can be changed as desired without
compromising ferroelectric properties. A recent report tested a related
idea, i.e. a thin layer of ZrO$_{2}$ as a buffer between Hf$_{0.5}$Zr$_{0.5}$O$_{2}$
and SiO$_{2}$/Si, and found that ferroelectricity in Hf$_{0.5}$Zr$_{0.5}$O$_{2}$
is significantly enhanced \citep{xiao2019performance}. This experimental
report, along with our theoretical predictions, should encourage experimental
researchers to pursue this idea in a variety of materials systems.

\section{Computational methods\label{sec:Methods5}}

\begin{figure*}
\begin{centering}
\includegraphics[width=0.75\textwidth]{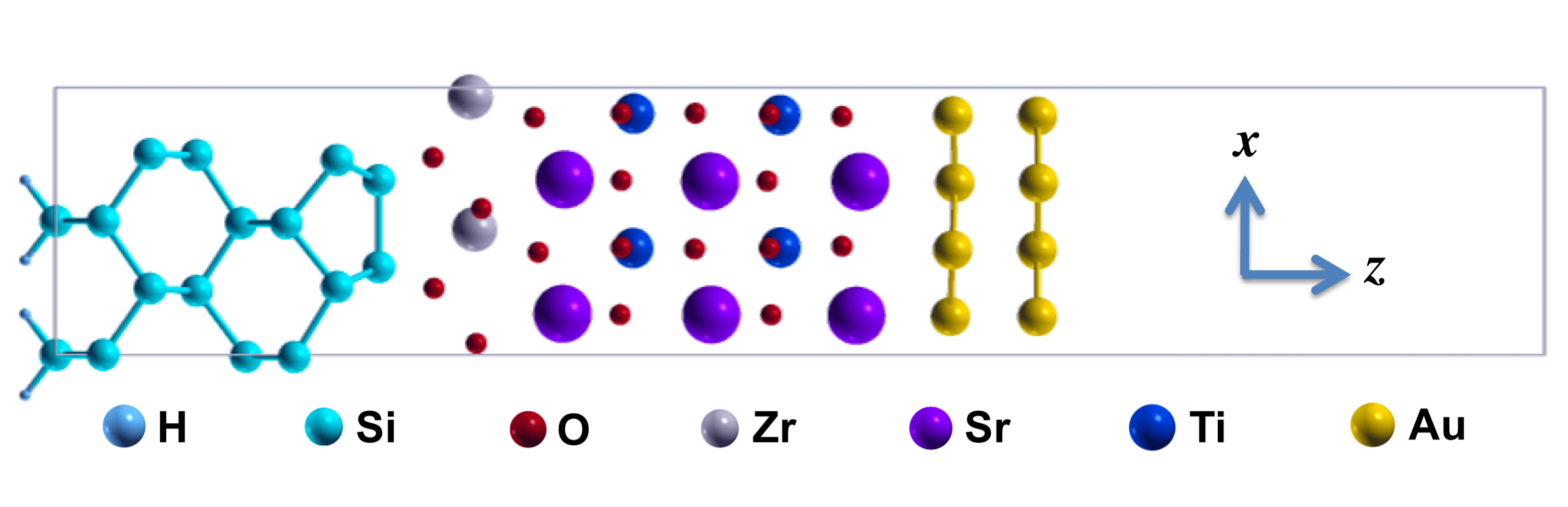}
\par\end{centering}
\caption{\label{fig:simcell}A typical simulation supercell of the SrTiO$_{3}$/ZrO$_{2}$/Si
stack with $2\times1$ in-plane periodicity. A two-layer Au top electrode
is also included in some simulations, and is shown. The bottom 4 layers
of Si are passivated by H and fixed to bulk coordinates. There is
$\sim12\text{\AA}$ of vacuum along the $z$-direction to separate
periodic copies.}
\end{figure*}

In order to find low-energy configurations of the systems of interest,
we use density functional theory (DFT) with the Perdew\--Burke\--Ernzerhof
generalized gradient approximation (PBE GGA) \citep{perdew1996generalized}
and ultrasoft pseudopotentials \citep{vanderbilt1990softselfconsistent}.
We employ the QUANTUM ESPRESSO software package \citep{giannozzi2009quantum}.
We use a $35$ Ry cutoff for the energy of the plane waves used to
describe the pseudo Kohn\--Sham wavefunctions. The Brillouin zone
is sampled with an $8\times8\times1$ Monkhorst\--Pack $k$-point
mesh (per $1\times1$ in-plane primitive cell) and a $0.02$ Ry Marzari\--Vanderbilt
smearing \citep{marzari1999thermal}. A typical simulation cell consists
of $8$ atomic layers of Si whose bottom layer is passivated with
H, a monolayer of ZrO$_{2}$, 1.5-3.5 unit cells of SrTiO$_{3}$,
and in some cases, 2 atomic layers of Au (see \figref{simcell}).
The in-plane lattice constant is fixed to $3.87\text{\AA}$, based
on the computed lattice constant of bulk silicon. $\sim12\text{\AA}$
of vacuum is placed between periodic copies of the slab in the $z$-direction.
However, the slab may have an overall dipole moment that might interact
with its periodic copies due to the long-range nature of the Coulomb
law. In order to eliminate this unphysical effect, a fictitious dipole
in the vacuum region of the cell is introduced so that it creates
an equal and opposite electric field in vacuum \citep{bengtsson1999dipolecorrection}.
All atomic coordinates are relaxed until the forces on all the atoms
are less than $10^{-3}\text{ Ry}/a_{0}$ in all axial directions,
where $a_{0}$ is the Bohr radius (the exception being the bottom
$4$ layers of Si which are fixed to their bulk positions in order
to simulate a thick Si substrate).

\section{Results \label{sec:Results5}}

\subsection{ZrO$_{x}$ monolayers on Si(001)\label{subsec:ZrOx-on-Si}}

The X-ray photoelectron spectroscopy (XPS) analysis presented in Ref.
\citep{dogan2018singleatomic} showed that in the ZrO$_{x}$/Si interface,
most of the interfacial Si atoms are in the Si$^{0}$ state, with
a small portion (< 0.2 monolayer) in the Si$^{1+}$ state. This established
that the growth procedure described in Ref. \citep{dogan2018singleatomic}
results in an interface without a large number of oxygen-coordinated
silicon atoms. \textcolor{black}{Further XPS analysis indicates that
most Zr atoms are in their Zr$^{4+}$ state with some in other oxidized
states.}\textcolor{red}{{} }Therefore, the experimental ZrO$_{x}$/Si
interface has $x\lesssim2$, and our previous theoretical work examined
the ZrO$_{2}$ stoichiometry \citep{dogan2019theoryof}. However,
we know that the oxygen content of oxide thin films can be highly
dependent on the growth conditions. In order to provide a comprehensive
survey of these monolayers that may be experimentally realized through
different growth methods, we have investigated ZrO$_{2}$ monolayers
with varying amounts of oxygen. To this end, we have simulated interfaces
with O:Zr ratios of 1.0, 1.5, 2.0, 2.5 and 3.0.

\subsubsection{Low-energy structures of ZrO$_{x}$ films}

In Ref. \citep{dogan2019theoryof}, we presented the low-energy structures
of ZrO$_{2}$ monolayers on the Si(001) surface. Among the several
metastable $2\times1$ configurations we discovered, five of them
are within 1 eV (per $2\times1$ cell) of the lowest-energy configuration.
We reported on the ionic polarizations, transition barriers and domain
energetics of these configurations in Ref. \citep{dogan2019theoryof}.
To generate the initial configurations of the under- and over-oxygenated
films, we take all the metastable configurations of the ZrO$_{2}$
monolayers, and either remove or add oxygens appropriately. For ZrO$_{x}$
where $x=1.0\left(1.5\right)$, we remove 2 (1) O per cell, which
yields 6 (4) initial configurations for each metastable configuration
of ZrO$_{2}$. For ZrO$_{x}$ where $x=2.5\text{ or }3.0$, we add
oxygen atoms to one or two of the following four positions: between
surface Si atoms along the $x$-direction (the midpoint of a dimer
and halfway between successive dimers for non-$1\times1$ systems),
and between a surface Si atom and its neighbor in the subsurface layer
(two inequivalent positions for non-$1\times1$ systems). These choices
are suggested by previous studies of O adsorption to the bare Si(001)
surface \citep{miyamoto1990atomicand,uchiyama1996atomicand}. We find
that the most favorable position for an O atom on the Si(001) surface
is the midpoint of a dimer, and the next most favorable position (higher
in energy by 0.23 eV per O) is between a surface atom and its subsurface
neighbor (a ``back bond''). Therefore, we have four positions to
add an oxygen to a $2\times1$ ZrO$_{2}$/Si interface, and hence
4 (6) initial configurations of ZrO$_{2.5\left(3.0\right)}$ for each
metastable configuration of ZrO$_{2}$.

\begin{figure}
\begin{centering}
\includegraphics[width=1\columnwidth]{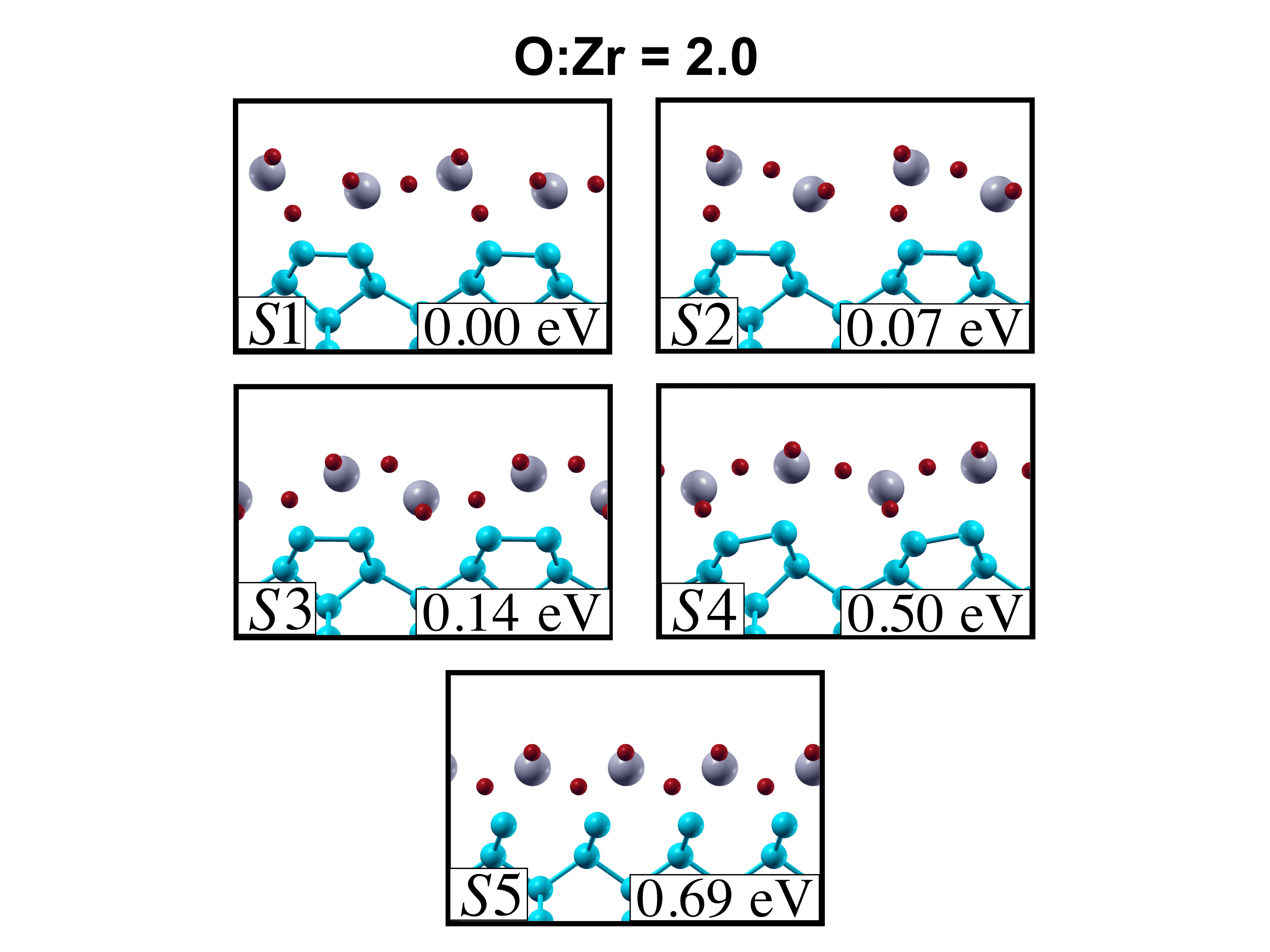}
\par\end{centering}
\caption{\label{fig:SiZrOx_20}Low energy relaxed configurations of monolayers
with O:Zr = 2.0. Energies are reported per $2\times1$ in-plane cell.}
\end{figure}

After relaxing all of the configurations we have generated according
to the above procedure, we have obtained a large number of metastable
configurations for ZrO$_{x}$ for $x=1.0,1.5,2.5\text{ and }3.0$.
In \tabref{SiZrOx_en} we list the energies of these configurations
(as well as ZrO$_{2}$), for structures that are 1 eV or less (per
$2\times1$ cell) higher than the ground state for all $x$. We follow
the naming convention in Ref. \citep{dogan2019theoryof}: for a given
stoichiometry, $S1$ corresponds to the ground state, $S2$ the second
lowest-energy state and so on. (Thus, two $S1$ structures for two
different stoichiometries are not structurally related.) We notice
that the multiplicity of structures at low energies is a feature of
this system independent of the oxygen content. For $x=1.0,2.5\text{ and }3.0$,
the second lowest energy structure is within $0.02$ eV of the ground
state. See \figref{SiZrOx_20} for the illustrations of the listed
structures for $x=2.0$, and the Supplementary Material for the other
values of $x$.

\begin{table}
\begin{centering}
\begin{tabular}{ccccccc}
\toprule 
\addlinespace[0.3cm]
$\ \ $$E$ (eV)$\ \ $ & $\ \ $$S1$$\ \ $ & $\ \ $$S2$$\ \ $ & $\ \ $$S3$$\ \ $ & $\ \ $$S4$$\ \ $ & $\ \ $$S5$$\ \ $ & $\ \ $$S6$$\ \ $\tabularnewline\addlinespace[0.3cm]
\addlinespace[0.1cm]
\midrule 
\addlinespace[0.3cm]
ZrO$_{1.0}$ & $\equiv$0.00 & 0.06 & 0.18 & 0.19 & 0.49 & 0.83\tabularnewline\addlinespace[0.3cm]
\addlinespace[0.1cm]
\midrule 
\addlinespace[0.3cm]
ZrO$_{1.5}$ & $\equiv$0.00 & 0.12 & 0.45 & 0.63 & 0.88 & \tabularnewline\addlinespace[0.3cm]
\addlinespace[0.1cm]
\midrule 
\addlinespace[0.3cm]
ZrO$_{2.0}$ & $\equiv$0.00 & 0.07 & 0.14 & 0.50 & 0.69 & \tabularnewline\addlinespace[0.3cm]
\addlinespace[0.1cm]
\midrule 
\addlinespace[0.3cm]
ZrO$_{2.5}$ & $\equiv$0.00 & 0.02 & 0.32 & 0.33 & 0.50 & 0.90\tabularnewline\addlinespace[0.3cm]
\addlinespace[0.1cm]
\midrule 
\addlinespace[0.3cm]
ZrO$_{3.0}$ & $\equiv$0.00 & 0.00 & 0.29 &  &  & \tabularnewline\addlinespace[0.3cm]
\addlinespace[0.1cm]
\end{tabular}
\par\end{centering}
\caption{\label{tab:SiZrOx_en}Total energies of ZrO$_{x}$ monolayers on Si(001),
where $x=1.0,1.5,2.0,2.5,3.0$ (reported in eV per $2\times1$ cell).}
\end{table}

\subsubsection{Polarization and Zr-O coordination}

Given the large number of structures, we are only able to describe
overall statistical trends for this dataset. Our main observations
are the following. (i) For the under-oxygenated cases ($x=1.0,1.5$),
we find that structures with lower energy have fewer Si-O bonds, and
on average the oxygens are farther from the Si surface than the Zr:
this represents the fact that O prefers to bond to Zr over Si. (ii)
For low oxygen content, we find more Si-Zr bonds forming: this happens
because the Zr are not fully oxidized and prefer to donate electrons
to the more electronegative Si. (iii) In the over-oxygenated cases
($x=2.5,3.0$), we find that the extra oxygens bond to the available
sites on the Si surface and the remaining oxygens distribute themselves
among the Zr atoms such that the coordination of the Zr by O is maximized.

\begin{figure}
\begin{centering}
\includegraphics[width=1\columnwidth]{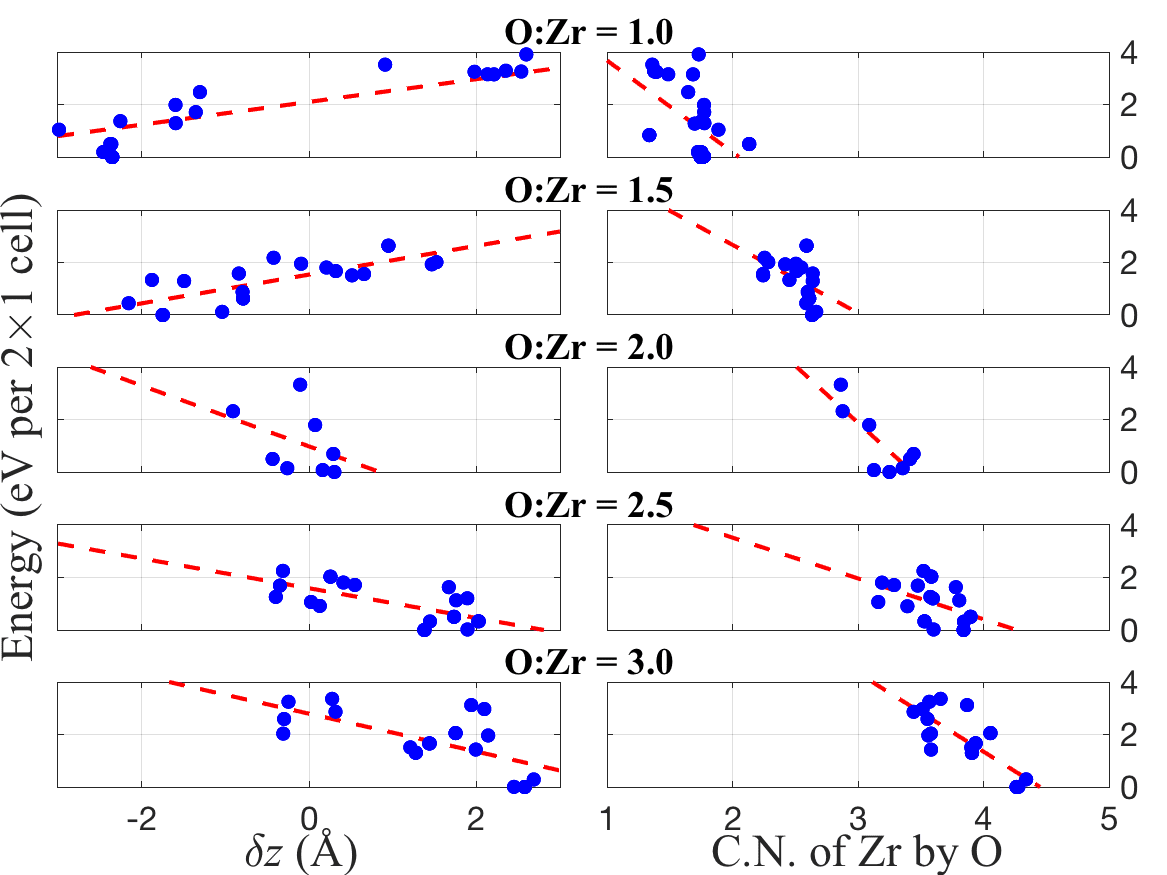}
\par\end{centering}
\caption{\label{fig:SiZrOx_Evsdz}Scatter plots of energy $E\ \text{versus}\ \delta z$
and the coordination number (C.N.) of Zr by O for the relaxed structures
of the ZrO$_{x}$ monolayers on Si(001), where $x=1.0,1.5,2.0,2.5,3.0$.
Linear least-squares fits are displayed as red dashed lines. The quantity
$\delta z\equiv\overline{z\left(\text{Zr}\right)}-\overline{z\left(\text{O}\right)}$
measures the mean ionic out-of-plane polarization. C. N. of Zr by
O is defined as the number of oxygen atoms that are within $\sim2.5\ \text{\AA}$
of a zirconium atom (see text for details). For each structure, the
C.N. is calculated separately for the two inequivalent Zr atoms and
then is averaged over the Zr. Energies are in eV per $2\times1$ in-plane
cell measured with respect to the lowest energy structure for each
O:Zr ratio.}
\end{figure}

The above observations follow from the data shown in \figref{SiZrOx_Evsdz}.
On the left hand side, we plot total energy vs $\delta z$ for all
of the structures in our library. The quantity $\delta z$ describes
the ionic polarization of the ZrO$_{x}$ monolayer and is defined
as the mean vertical Zr-O separation: $\delta z\equiv\overline{z\left(\text{Zr}\right)}-\overline{z\left(\text{O}\right)}$.
A crude interpretation of this quantity suggests that the film is
positively polarized when $\delta z>0$ and vice versa. We find that
negative ionic polarization is preferred for $x=1.0,1.5$, and that
the energy increases as the polarization increases. Hence, the O anions
prefer to be farther from the Si surface, and the Zr cations prefer
to be closer. We find the opposite for $x=2.0,2.5,3.0$, where positive
polarization is preferred, and the lowest energy structures have the
highest Zr-O out-of-plane separation. On the right hand side, we plot
total energy vs the coordination number of Zr by O. We calculate the
coordination number for a Zr atom by counting the number of O atoms
within $\sim2.5\ \text{\AA}$, which is approximately the sum of the
Zr and O atomic radii. Instead of a sharp cutoff at $2.5\ \text{\AA}$,
for each Zr-O bond, we use a Fermi\--Dirac function centered at $2.7\ \text{\AA}$
(``chemical potential'') and with width $0.4\ \text{Å}$ (``temperature'')
to compute coordination numbers. We then average the coordination
numbers of the two inequivalent Zr atoms and report it in the figure.
We find that for all O:Zr ratios, higher C.N. correlates with lower
energy.

The above analysis of the ZrO$_{x}$ films on Si with varying O:Zr
ratios suggests that the oxygen content can be used to change the
ferroelectric switching behavior of the films. Indeed, in the case
of ZrO$_{2.0}$, our in-depth investigation of domain energetics indicated
that the polarization switching likely occurs between $S2$ and $S3$
(see \tabref{SiZrOx_en}) \citep{dogan2019theoryof}, which results
in an energy difference of 0.07 eV and a polarization difference (change
in $\delta z$) of $0.42\ \text{\AA}$. We list the analogous values
for all low-energy transitions in \tabref{SiZrOx_transitions}. For
example, for ZrO$_{1.5}$ (ZrO$_{2.5}$), if the switching occurred
between $S1$ and $S2$, the energy difference would be 0.12 eV (0.02
eV) and the polarization difference would be $0.71\ \text{\AA}$ ($0.51\ \text{\AA}$).
In the table, we only include transitions between low-energy con\textcolor{black}{figurations
($E_{\textrm{initial}},E_{\textrm{final}}\leq0.20$ eV with respect
to $S1$) with a low energy difference ($\left|E_{\textrm{final}}-E_{\textrm{initial}}\right|\leq0.15$)}.
We observe that for $x=1,5,2.0$ and 2.5, there are low-energy transitions
with large changes in the ionic polarization, and thus, they may be
the best compositions for ferroelectric applications. It may be worth
testing experimentally whether there is a larger ferroelectric switching
in the case of $x=1.5$ and 2.5 compared to $x=2.0$, as suggested
by our results.

\begin{table}
\begin{centering}
\begin{tabular}{cccc}
\toprule 
\addlinespace[0.3cm]
 & Transition & Change in $E$ (eV) & Change in $\delta z\ \left(\text{\AA}\right)$\tabularnewline\addlinespace[0.3cm]
\addlinespace[0.1cm]
\midrule 
\addlinespace[0.3cm]
\multirow{4}{*}{ZrO$_{1.0}$} & $S1\rightarrow S2$ & 0.06 & -0.08\tabularnewline\addlinespace[0.3cm]
\cmidrule{2-4} 
\addlinespace[0.1cm]
\addlinespace[0.3cm]
 & $S2\rightarrow S3$ & 0.12 & 0.02\tabularnewline\addlinespace[0.3cm]
\cmidrule{2-4} 
\addlinespace[0.1cm]
\addlinespace[0.3cm]
 & $S3\rightarrow S4$ & 0.01 & 0.03\tabularnewline\addlinespace[0.3cm]
\cmidrule{2-4} 
\addlinespace[0.1cm]
\addlinespace[0.3cm]
 & $S2\rightarrow S4$ & 0.13 & 0.05\tabularnewline\addlinespace[0.3cm]
\addlinespace[0.1cm]
\midrule 
\addlinespace[0.3cm]
ZrO$_{1.5}$ & $S1\rightarrow S2$ & 0.12 & 0.71\tabularnewline\addlinespace[0.3cm]
\addlinespace[0.1cm]
\midrule 
\addlinespace[0.3cm]
\multirow{3}{*}{ZrO$_{2.0}$} & $S1\rightarrow S2$ & 0.07 & -0.15\tabularnewline\addlinespace[0.3cm]
\cmidrule{2-4} 
\addlinespace[0.1cm]
\addlinespace[0.3cm]
 & $S2\rightarrow S3$ & 0.07 & -0.42\tabularnewline\addlinespace[0.3cm]
\cmidrule{2-4} 
\addlinespace[0.1cm]
\addlinespace[0.3cm]
 & $S1\rightarrow S3$ & 0.14 & -0.57\tabularnewline\addlinespace[0.3cm]
\addlinespace[0.1cm]
\midrule 
\addlinespace[0.3cm]
ZrO$_{2.5}$ & $S1\rightarrow S2$ & 0.02 & -0.51\tabularnewline\addlinespace[0.3cm]
\addlinespace[0.1cm]
\midrule 
\addlinespace[0.3cm]
ZrO$_{3.0}$ & $S1\rightarrow S2$ & 0.00 & 0.13\tabularnewline\addlinespace[0.3cm]
\addlinespace[0.1cm]
\end{tabular}
\par\end{centering}
\caption{\label{tab:SiZrOx_transitions}List of low-energy transitions for
ZrO$_{x}$ monolayers on Si(001), where $x=1.0,1.5,2.0,2.5,3.0$,
the change in energy (eV per $2\times1$ cell), and the change in
the average cation-anion vertical displacement ($\text{\AA}$) as
a result of the transition.}
\end{table}

\subsubsection{Electronic structure of ZrO$_{x}$ films}

In many potential applications of the ZrO$_{x}$/Si interface such
as the FEFET, an insulating interface is desired. In \tabref{SiZrOx_gap},
we list the computed band gaps of the low energy configurations of
the ZrO$_{x}$/Si stacks. The maximal value of 1.0 eV is equal to
the band gap of Si in the interior of the substrate, as determined
by an analysis of layer-by-layer projected densities of states. We
compute the band gap of bulk silicon as 0.7 eV, which is smaller than
the experimental gap of 1.2 eV \citep{kittel2004introduction}, but
in agreement with other computational studies employing GGA \citep{heyd2005energyband}.
This underestimation of the gap is expected in DFT. However, in the
8-layer thick Si slab we have used, the gap increases to 1.0 eV due
to quantum confinement. We have tested this effect by varying the
slab thickness in Si thin films: for 8, 12, 16 and 20 layers, we have
found a band gap of 1.02, 0.94, 0.85 and 0.78 eV, respectively.

According to \tabref{SiZrOx_gap}, for $x=1.5,2.0$ and 2.5, the low-energy
configurations are insulating with the maximal (or close to the maximal)
band gap. This reinforces the usefulness of these compositions in
FEFET-related applications. We observe that for ZrO$_{1.0}$, the
two lowest-energy configurations have a small gap and the higher-energy
configurations are metallic. This is due to the under-coordination
of zirconiums by oxygens and/or silicon dangling bonds. This is also
the case for $S5$ of ZrO$_{1.5}$. Therefore a lower O:Zr ratio should
be aimed for other applications in which a metallic interface is desired.

\begin{table}
\begin{centering}
\begin{tabular}{ccccccc}
\toprule 
\addlinespace[0.3cm]
$\ \ $$E_{\text{gap}}$ (eV)$\ \ $ & $\ \ $$S1$$\ \ $ & $\ \ $$S2$$\ \ $ & $\ \ $$S3$$\ \ $ & $\ \ $$S4$$\ \ $ & $\ \ $$S5$$\ \ $ & $\ \ $$S6$$\ \ $\tabularnewline\addlinespace[0.3cm]
\addlinespace[0.1cm]
\midrule 
\addlinespace[0.3cm]
ZrO$_{1.0}$ & 0.3 & 0.2 & metal & metal & metal & metal\tabularnewline\addlinespace[0.3cm]
\addlinespace[0.1cm]
\midrule 
\addlinespace[0.3cm]
ZrO$_{1.5}$ & 1.0 & 1.0 & 0.7 & 0.7 & metal & \tabularnewline\addlinespace[0.3cm]
\addlinespace[0.1cm]
\midrule 
\addlinespace[0.3cm]
ZrO$_{2.0}$ & 1.0 & 1.0 & 1.0 & 0.6 & 1.0 & \tabularnewline\addlinespace[0.3cm]
\addlinespace[0.1cm]
\midrule 
\addlinespace[0.3cm]
ZrO$_{2.5}$ & 0.8 & 0.9 & 0.9 & 0.8 & 0.9 & 1.0\tabularnewline\addlinespace[0.3cm]
\addlinespace[0.1cm]
\midrule 
\addlinespace[0.3cm]
ZrO$_{3.0}$ & 0.7 & 0.7 & 0.7 &  &  & \tabularnewline\addlinespace[0.3cm]
\addlinespace[0.1cm]
\end{tabular}
\par\end{centering}
\caption{\label{tab:SiZrOx_gap}DFT band gaps of ZrO$_{x}$ monolayers on Si(001),
where $x=1.0,1.5,2.0,2.5,3.0$ (reported in eV).}
\end{table}

\subsection{ZrO$_{2}$ as a buffer between SrTiO$_{3}$ and Si\label{subsec:STO_ZrOx}}

The observation of ferroelectricity in the monolayer ZrO$_{2}$ on
silicon marks the experimental attainment of the thinnest possible
oxide ferroelectric \citep{dogan2018singleatomic,dogan2019theoryof}.
In the previous section, we have shown that under- and over-oxygenated
ZrO$_{2}$ may also exhibit switchable polarization. A related goal
in the field of thin film oxide/semiconductor physics is to achieve
ferroelectricity in thin perovskite films \citep{dogan2017abinitio}.
To this end, we have conducted a study of SrTiO$_{3}$/ZrO$_{2}$/Si
heterostructures. Using MBE, SrTiO$_{3}$ can be grown epitaxially
on Si with a small compressive strain \citep{mckee1998crystalline,kolpak2010interfaceinduced}.
It has also been found that although SrTiO$_{3}$ is not a ferroelectric
material in the bulk down to very low temperatures, a small compressive
strain, attainable by epitaxy to the Si(001) surface, makes it a room-temperature
ferroelectric \citep{haeni2004roomtemperature}. However, the SrTiO$_{3}$/Si
heterostructures grown to date have been paraelectric due to the pinning
of the polarization by the interface chemistry \citep{kolpak2010interfaceinduced,kolpak2012interface}.
We will show below that it is possible to overcome this pinning with
a buffer layer that has a richer landscape of interface chemistry,
i.e. ZrO$_{2}$, and that STO/ZrO$_{2}$/Si heterostructures indeed
possess multiple (meta)stable configurations with varying polarization
profiles. This may enable researchers to grow oxide/semiconductor
heterostructures of varying thicknesses with switchable polarization.

\subsubsection{Low-energy configurations of SrTiO$_{3}$/ZrO$_{2}$/Si stacks}

\begin{figure*}
\begin{centering}
\includegraphics[width=0.8\textwidth]{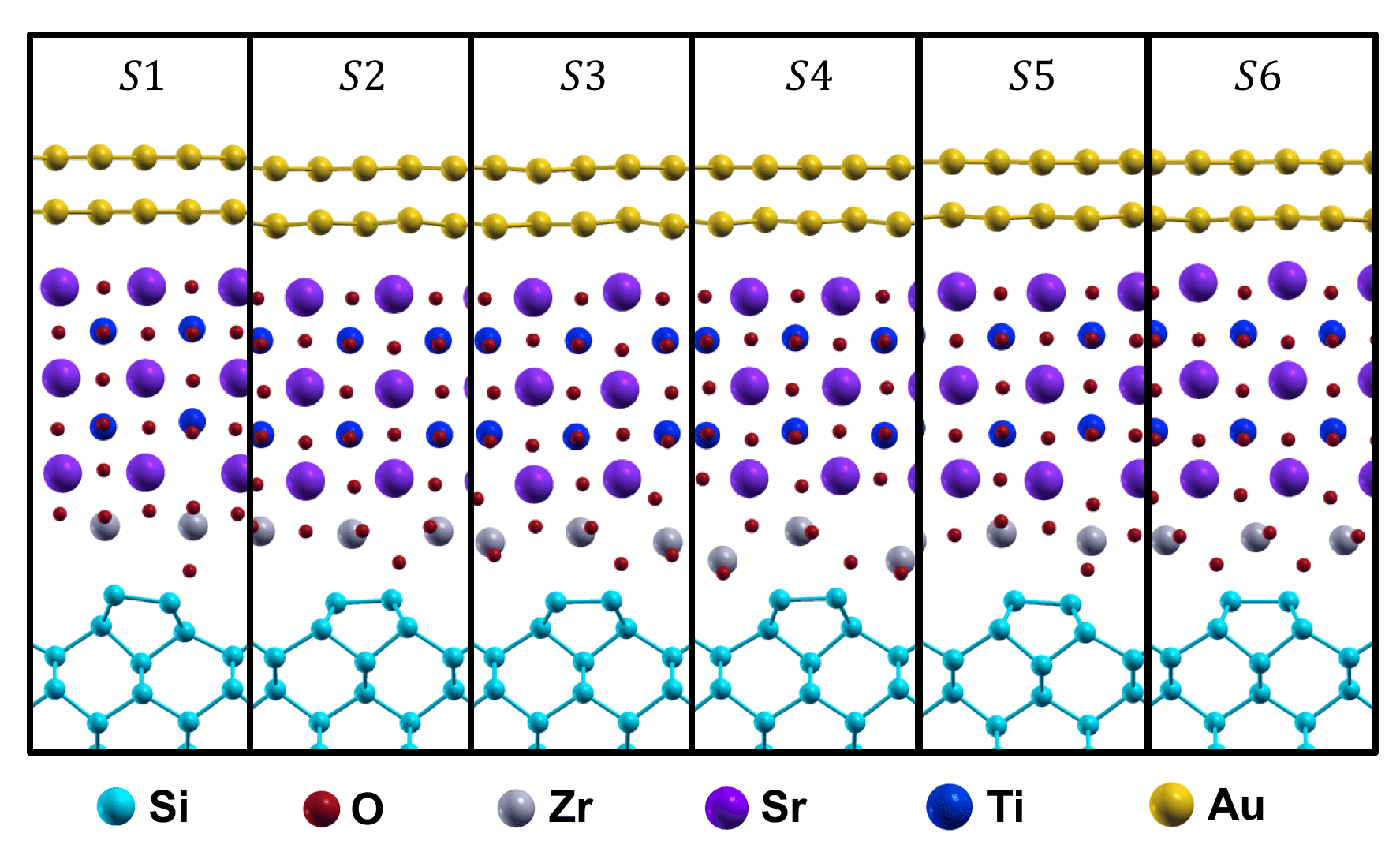}
\par\end{centering}
\caption{\label{fig:SiSTO_str}Relaxed SrTiO$_{3}$/ZrO$_{2}$/Si heterostructures
with 2.5 u.c.-thick STO and Au top electrode. For each configuration,
the displayed portion of the stack is slightly larger than the $2\times1$
unit cell.}
\end{figure*}

We find that bulk SrTiO$_{3}$ has a lattice constant of $3.93\text{ \AA}$,
which puts it at a $1.5\%$ compressive strain on Si(001), in agreement
with previous studies \citep{kolpak2010interfaceinduced,kolpak2012interface}.
In order to generate initial configurations for STO/ZrO$_{2}$/Si
stacks, we have begun with the relaxed coordinates of the five low-energy
ZrO$_{2}$/Si interfaces, and placed a 1.5 u.c.-thick STO slab on
top, laterally shifted by the vector $\left(\frac{n_{1}}{3}a,\frac{n_{2}}{3}a,0\right)$,
where $n_{1},n_{2}=0,1,2$, and $a$ is the lattice constant, which
yields 45 initial configurations. Relaxing these 45 configurations
have resulted in 6 configurations which are all local minima in the
energy landscape. We have then added more STO layers to generate 2,
2.5, 3 and 3.5 u.c.-thick slabs. Finally, to create a gate stack and
to examine the effects of boundary conditions, we have added a two-layer
thick gold top electrode to all of these relaxed configurations. In
total we generated 6 (interfaces) $\times$ 5 (thicknesses) $\times$
2 (with and without the top electrode) $=$ 60 configurations. The
2.5 u.c.-thick slabs with the gold electrode are displayed in \figref{SiSTO_str}.

The immediate observations from \figref{SiSTO_str} are that (i) all
the interfaces have dimerized silicon, preserving the $2\times1$
periodicity of the Si(001) surface, and (ii) in 2 interfaces, there
is migration of oxygen from the first SrO layer to the ZrO$_{2}$
layer (a full migration in $S1$ and sharing of an oxygen between
the two layers in $S5$). For all configurations, from the first (bottom)
TiO$_{2}$ layer up to the top layer, STO possesses the stoichiometric
perovskite structure. Therefore, while examining the polarization
profile of these stacks, we start from the first TiO$_{2}$ layer.

\subsubsection{Interfacial chemistry}

\begin{figure*}
\begin{centering}
\includegraphics[width=0.8\textwidth]{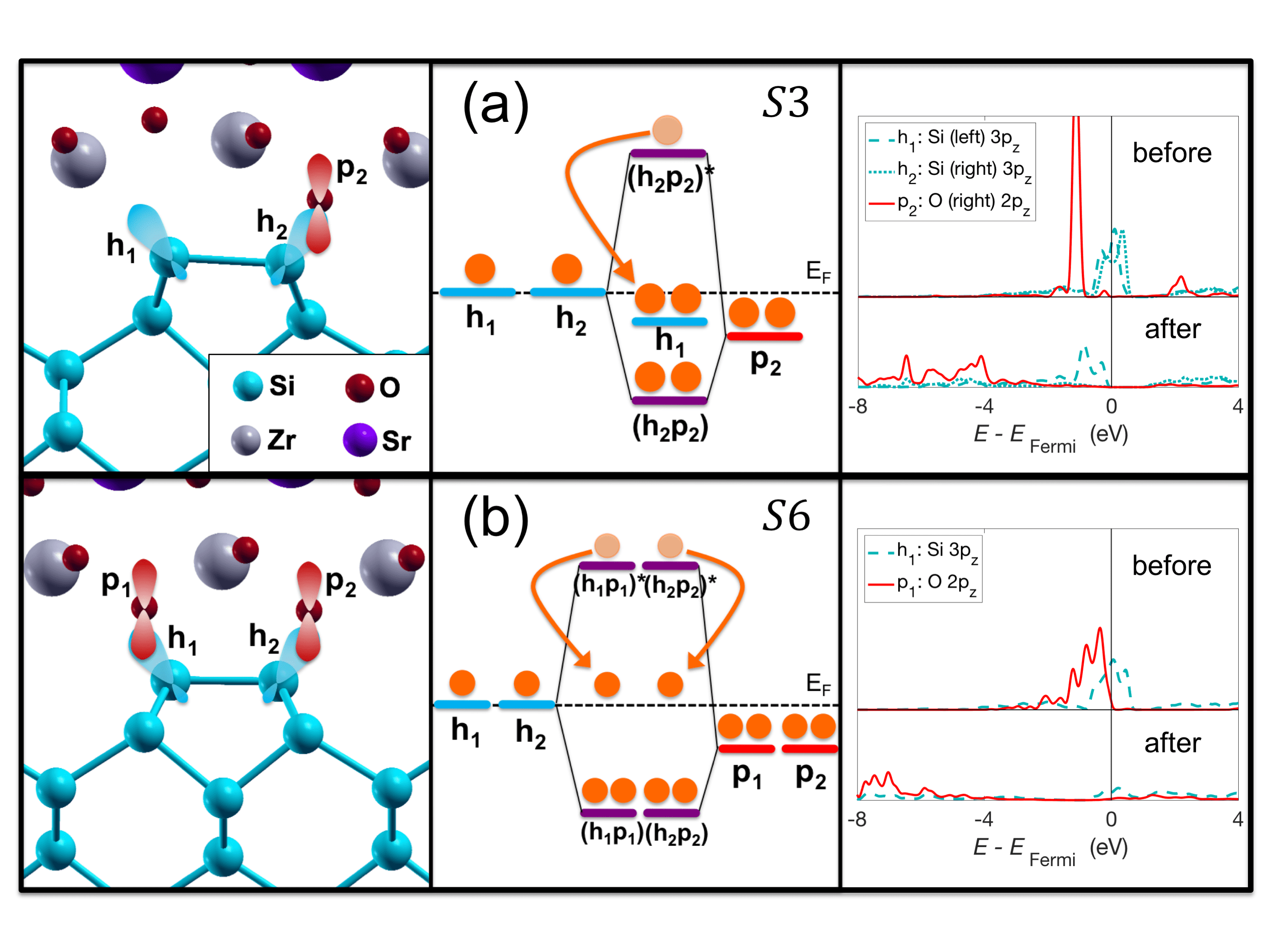}
\par\end{centering}
\caption{\label{fig:SiSTO_chem}Chemical bonding at the ZrO$_{2}$/Si interface
for SrTiO$_{3}$/ZrO$_{2}$/Si stacks in the configurations (a) $S3$
and (b) $S6$. For each configuration, the atomic structure in the
vicinity of this interface, as well as the schematics that represent
the atomic orbitals participating in the chemical bonding are displayed
in the left panel (see text for details); a level diagram that summarizes
the simplified chemical bonding model of the interface is displayed
in the middle panel; and densities of states projected onto the relevant
atomic orbitals (PDOS) before and after the interface is formed are
displayed in the right panel (only $h_{1}$ and $p_{1}$ are shown
for $S6$ because of their approximate equivalence to $h_{2}$ and
$p_{2}$ in this configuration). For the PDOS plots, the zero of the
energy is taken as the Fermi level.}
\end{figure*}

In an oxide/semiconductor interface such as ZrO$_{2}$/Si, the chemical
bonding at the interface is expected to determine the electronic structure
of the stack, which then influences the polarization profile \citep{kolpak2012interface,dogan2017abinitio}.
A simple inspection of \figref{SiSTO_str} suggests that there are
two types of ZrO$_{2}$/Si interfaces present: (1) where one of the
atoms in the silicon dimer bonds with an oxygen ($S1$ through $S5$),
and (2) where both of the atoms in the silicon dimer bond with oxygens
($S6$). In both types, these bonds are between the dangling hybrid
orbitals of Si (with $sp^{3}$ character) and the $2p_{z}$ orbitals
of interfacial O. In \figref{SiSTO_chem}, we describe the interfacial
chemistry for both types and use $S3$ to represent the first type.
The Si dangling hybrid orbitals are labeled $h_{1}$ and $h_{2}$,
and the participating oxygen $p_{z}$ orbitals are labeled $p_{1}$
and $p_{2}$. In the left panel, the interface geometries for (a)
$S3$ and (b) $S6$ are displayed, with the schematics of participating
orbitals overlaid on their respective atoms (for $S3$, $p_{1}$ does
not significantly with $h_{1}$, and hence does not participate).

For $S3$, prior to the formation of the interface, the oxygen atoms
are in the $\text{O}^{2-}$ state, thus $p_{2}$ starts out with two
electrons, whereas $h_{1}$ and $h_{2}$ have one electron each (middle
panel in \figref{SiSTO_chem}(a)). Once the interface is formed, two
of the three electrons in the $h_{2}$ and $p_{2}$ orbitals occupy
the $\left(h_{2}p_{2}\right)$ bonding states, and the remaining electron
is accepted by $h_{1}$. In the right panel of \figref{SiSTO_chem}(a),
we display the projected densities of states (PDOS) of these orbitals,
before and after the interface is formed. We use Si $3p_{z}$ orbitals
to approximate $h_{1}$ and $h_{2}$ since we expect the these orbitals
to be in close alignment with the $z$-axis for a dimerized surface.
In the figure, we see that $h_{1}$ and $h_{2}$ are half-occupied
before the interface is formed, and $p_{2}$ is fully occupied. After
the interface is formed, $h_{2}$ mixes strongly with $p_{2}$ and
their spectral features become broad at low energies while the PDOS
for $h_{1}$ becomes fully occupied without a significant change in
its shape.

For $S6$, in contrast to $S3$, both $p_{1}$ and $p_{2}$ participate
in the chemical bonding in the same way. After the formation of the
$\left(h_{1}p_{1}\right)$ and $\left(h_{2}p_{2}\right)$ bonds, an
electron per bond dopes the Fermi level, due to the absence of available
unoccupied or partially occupied interface states. In the right panel
of \figref{SiSTO_chem}(b), PDOS for $h_{1}$ and $p_{1}$ are displayed
before and after the interface is formed ($h_{2}$ and $p_{2}$ are
almost identical to their counterparts, and hence not shown)\textcolor{black}{.
Upon the formation of the interface, $h_{1}$ and $p_{1}$ mix, the
electrons are donated to the Fermi level, and the Fermi level enters
the conduction band (see DOS of Figure \ref{fig:SiSTO_DOS}(b)) indicating
electron doping into the conduction band.}

\begin{figure}
\begin{centering}
\includegraphics[width=0.8\columnwidth]{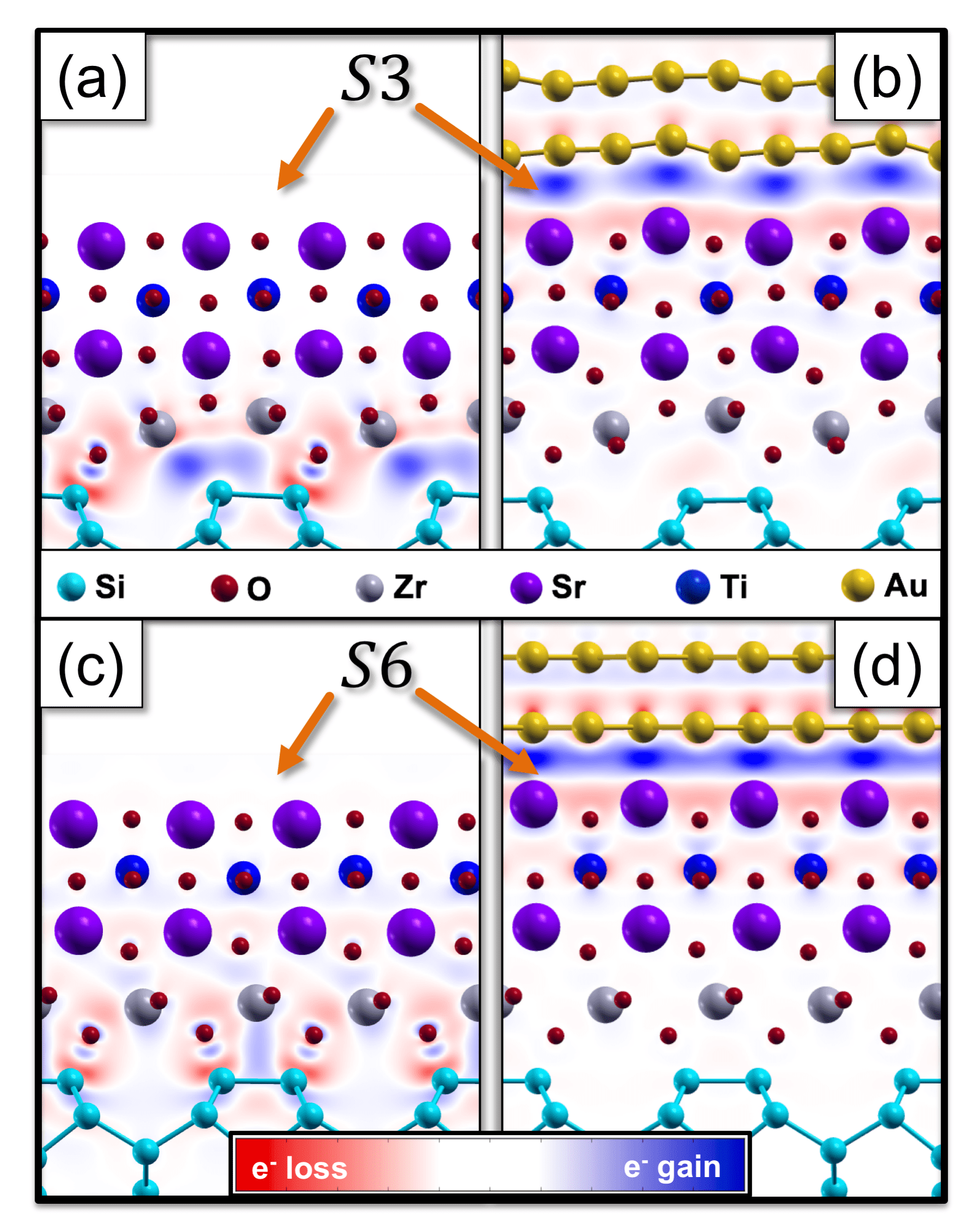}
\par\end{centering}
\caption{\label{fig:SiSTO_ecd}Electron density redistribution for (1.5 u.c.
SrTiO$_{3}$)/ZrO$_{2}$/Si heterostructures upon the formation of
the ZrO$_{2}$/Si interface for the (a) $S3$ and (c) $S6$ configurations;
and upon the addition of the capping electrode for the (b) $S3$ and
(d) $S6$ configurations. The plots are obtained by averaging the
electron density redistribution along the $1\times$ direction.}
\end{figure}

To further support our simple picture of the interfacial chemistry,
we have computed the electron redistribution for (1.5 u.c. SrTiO$_{3}$)/ZrO$_{2}$/Si
stacks, defined as $\Delta n\left(x,y,z\right)=n_{\text{STO/ZrO}_{2}\text{/Si}}\left(x,y,z\right)-n_{\text{STO/ZrO}_{2}}\left(x,y,z\right)-n_{\text{Si}}\left(x,y,z\right).$
We then average over the $1\times$ direction and obtain $\overline{\Delta n}\left(x,z\right)$,
which we display in \figref{SiSTO_ecd}(a) and \figref{SiSTO_ecd}(c)
for $S3$ and $S6$, respectively. For $S3$, the figure indicates
that the electron density around the Si-O $\left(h_{2}p_{2}\right)$
bond decreases while the density in the region between the un-bonded
Si atom and the neighboring Zr atom increases. For $S6$, on the other
hand, the electron density around the Si-O bonds decreases while the
regions of increasing electron density are spatially distributed in
the oxide. Plots analogous to \figref{SiSTO_ecd} for the remaining
configurations ($S1$, $S2$, $S4$ and $S5$) are presented in the
Supplementary Material. The simple interfacial chemistry model we
have obtained for $S3$ also apply to these configurations. 

\begin{figure}
\begin{centering}
\includegraphics[width=1\columnwidth]{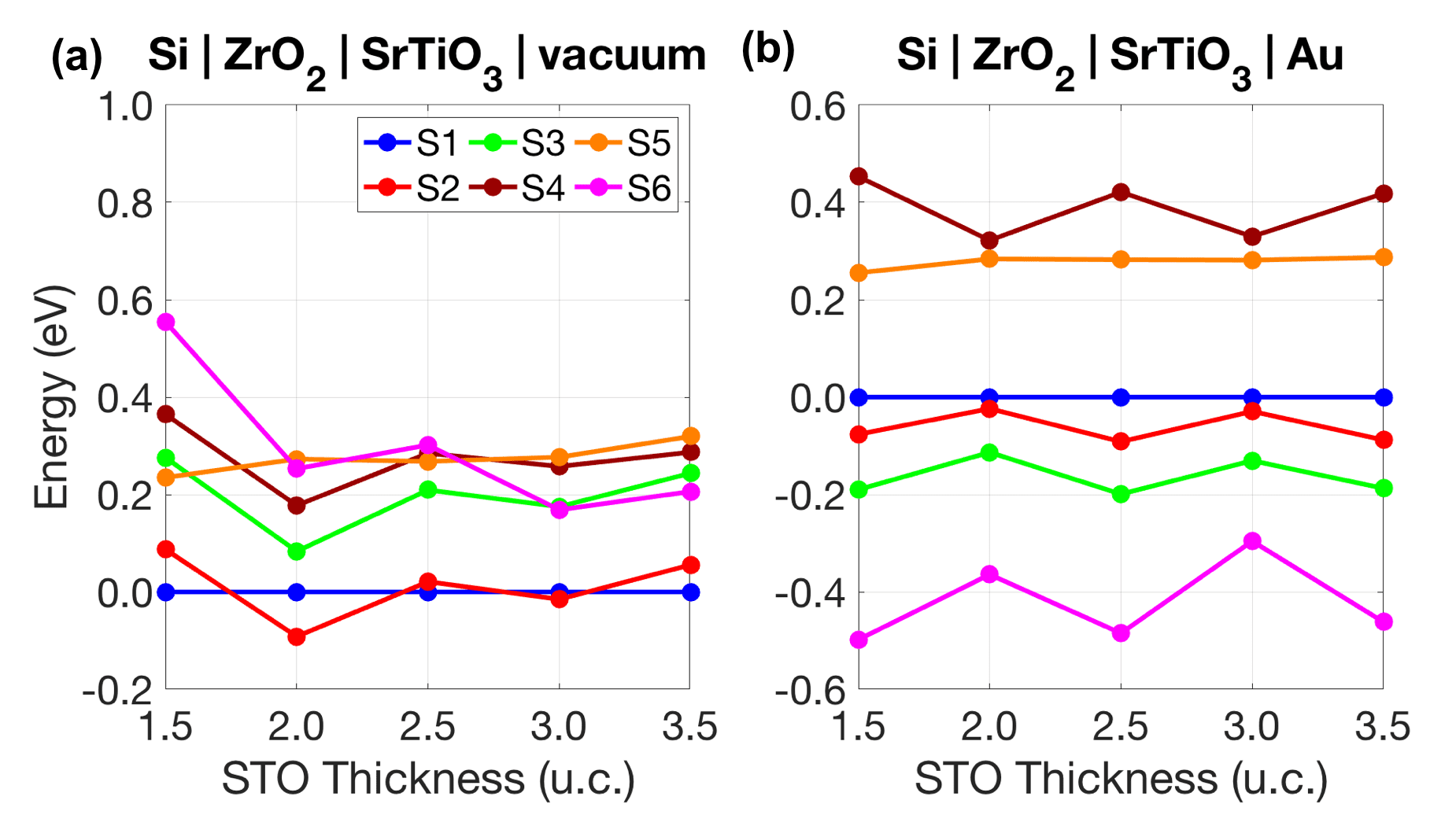}
\par\end{centering}
\caption{\label{fig:SiSTO_energy}Energies of the SrTiO$_{3}$/ZrO$_{2}$/Si
heterostructures with respect to the $S1$ configuration vs STO thickness,
(a) without a top electrode, and (b) with Au as the top electrode.
The energy is given in eV per $2\times1$ unit cell.}
\end{figure}

We present the total energies of the considered films in \figref{SiSTO_energy}.
The total energy of the $S1$ configuration is taken as the reference
for each case. We make two observations: (i) relative energies change
with film thickness but generally stay within $\pm0.1$ eV once the
STO thickness is above 2 u.c.; and (ii) the energy ordering is significantly
affected by the inclusion of the gold electrode. The largest changes
in relative energy with the addition of the electrode are for $S6$
(approx. -0.7 eV) and $S3$ (approx. -0.4 eV). These most stabilized
interfaces are also the ones with the largest polarization enhancement
when the electrode is added (\figref{SiSTO_profile} and \figref{SiSTO_averpol}).
Therefore, a subset of the possible interfaces become especially stabilized
by a high-work-function electrode such as gold. As we will discuss
below, this stabilization occurs by an electron transfer to the electrode
which was previously observed in the BaTiO$_{3}$/Ge system \citep{dogan2017abinitio}.
With the usage of the electrode with a finely tuned work function,
the lowest energy interfaces ($S6$ and $S3$ in this case) can in
principle be made degenerate, enabling polarization switching without
an energy cost.

\subsubsection{Film polarization}

\begin{figure}
\begin{centering}
\includegraphics[width=1\columnwidth]{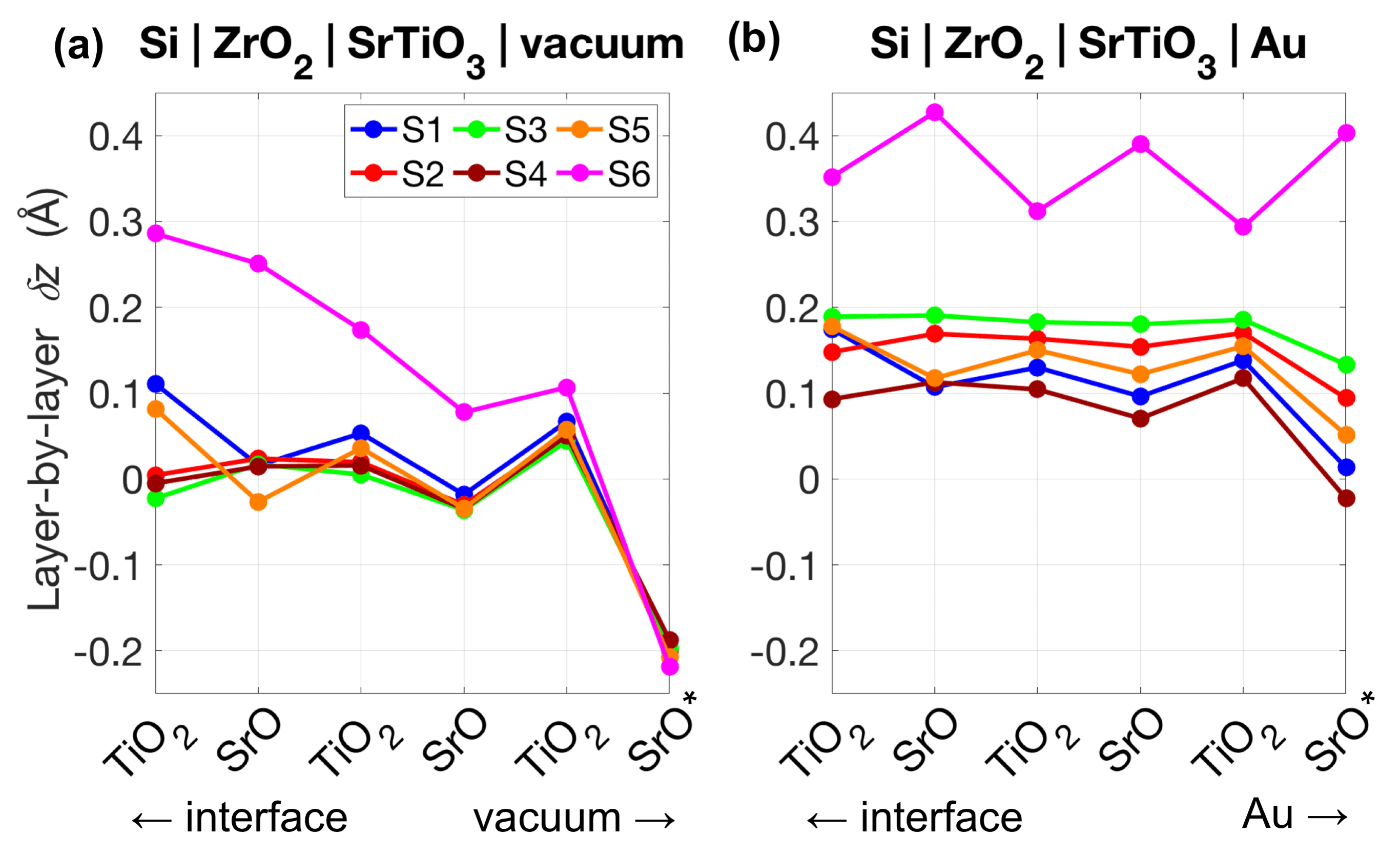}
\par\end{centering}
\caption{\label{fig:SiSTO_profile}Layer-by-layer polarization profile, measured
by mean vertical cation-anion separation ($\delta z$) within a layer
for 3.5 u.c.-thick STO films in SrTiO$_{3}$/ZrO$_{2}$/Si heterostructures,
(a) without a top electrode, and (b) with Au as the top electrode.
The top SrO layer (farthest from the interface) is marked with an
asterisk for both cases.}
\end{figure}

In order to assess the possibility of ferroelectric switching in the
SrTiO$_{3}$/ZrO$_{2}$/Si stacks, we have computed the layer-by-layer
mean vertical cation-anion separation: $\delta z\equiv\overline{z\left(\text{Sr or Ti}\right)}-\overline{z\left(\text{O}\right)}$
where the averaging is done over a layer. We present the polarization
profile of 3.5 u.c.-thick SrTiO$_{3}$ films in \figref{SiSTO_profile},
starting with the bottom TiO$_{2}$ layer. We observe that for the
films without a top electrode (\figref{SiSTO_profile}(a)), the polarization
at the surface SrO layer is pinned to the same value for all interface
configurations, shrinking the variation of the polarization among
the configurations in the upper layers. We also see that generally
the $\delta z$ values near the ZrO$_{2}$ are larger and quickly
decay toward the middle STO layers. This is due to the depolarizing
field not being screened by mobile charges or a capping electrode.
When the electrode is added, the depolarizing field is screened by
the metal, the top SrO polarization is no longer pinned, and thus
varies with the interface configuration (\figref{SiSTO_profile}(b)).
Furthermore, the polarization values for each interface increases
with the capping electrode. This can be explained by a high-work-function
electrode such as Au pulling mobile electrons from the STO/ZrO$_{2}$/Si
system, thereby creating an electric field that attracts cations and
repels anions. This effect is the largest in $S6$, which also has
the largest polarization values for non-capped systems in \figref{SiSTO_profile}(a),
indicating that these interfaces have more mobile carriers in them.
This is in agreement with our discussion above regarding the interfacial
chemistry of these configurations. These mobile carriers both (a)
screen the depolarizing field in the absence of an electrode, causing
the film to have larger polarization values, and (b) migrate to the
capping electrode, further enhancing the ionic polarization.

\begin{figure}
\begin{centering}
\includegraphics[width=1\columnwidth]{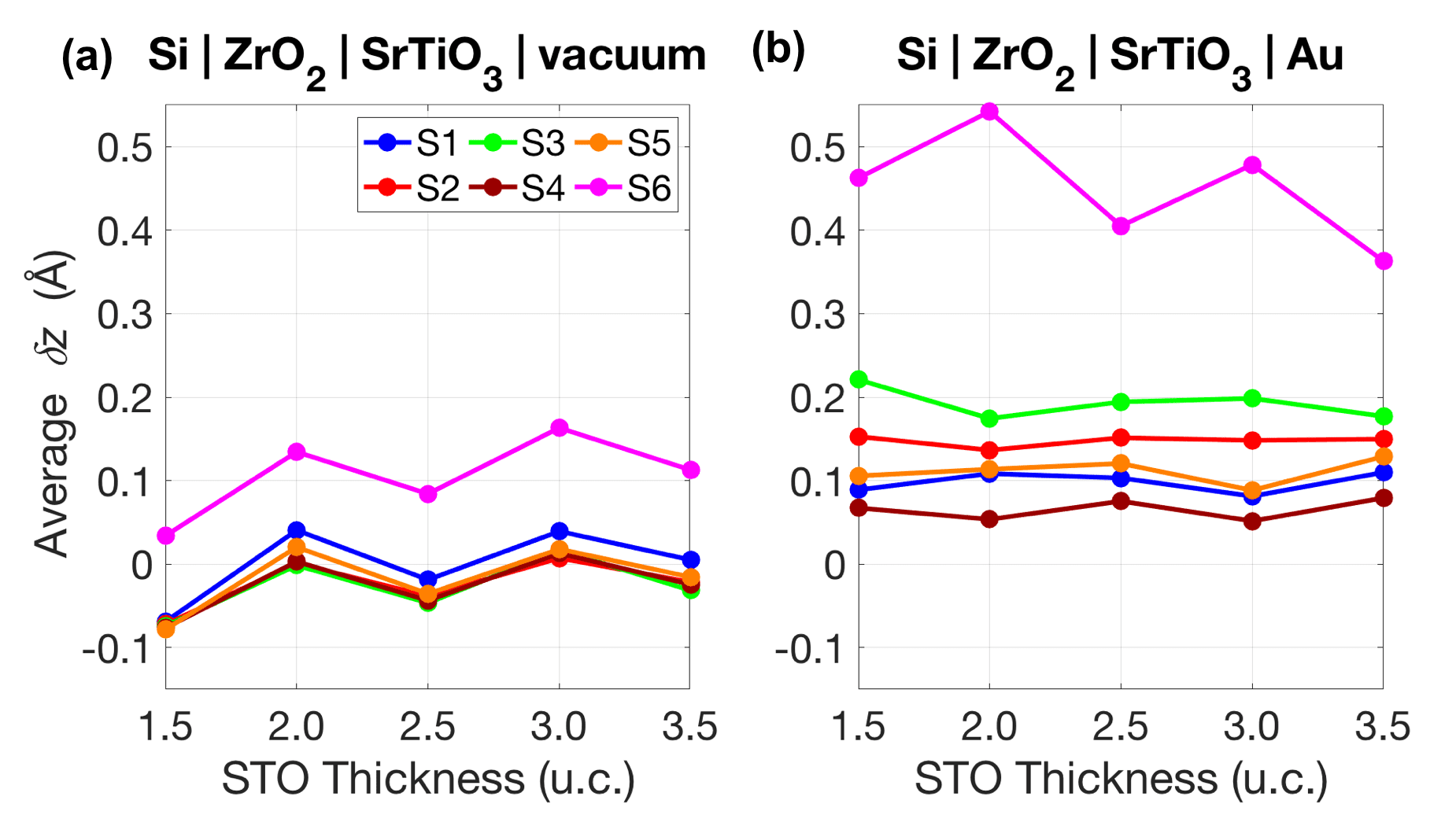}
\par\end{centering}
\caption{\label{fig:SiSTO_averpol}Average polarization, measured by taking
the average of $\delta z$ for all the layers in a given film, vs
STO thickness in SrTiO$_{3}$/ZrO$_{2}$/Si heterostructures, (a)
without a top electrode, and (b) with Au as the top electrode.}
\end{figure}

We report the average polarization for all the films we have computed
in \figref{SiSTO_averpol}. A universal feature of these results is
that the oxide's surface termination causes a small modulation in
the average polarization, due to the pinning of the surface polarization
to different values. For the films with no electrode (\figref{SiSTO_averpol}(a)),
the average polarization is close to zero, and does not significantly
change with thickness, with the exception of $S6$. Regarding $S6$,
because it has positive polarization values in the interior of STO
but a negative value at the pinned surface layer, its average polarization
value increases with increased thickness. As for the films with the
capping electrode (\figref{SiSTO_averpol}(b)), the charge transfer
to the electrode causes all structures to have positive average polarizations
that do not significantly depend on the thickness, again with the
exception of $S6$. For this configuration, there is a slight downward
slope for the average polarization when the thickness is increased.
Our findings on polarization profiles and electrode effects are in
line with previous studies on SrTiO$_{3}$/Si \citep{kolpak2012interface}
and BaTiO$_{3}$/Ge interfaces \citep{dogan2017abinitio}.

A comparison of the polarization values with and without the gold
electrode in \figref{SiSTO_profile} and \figref{SiSTO_averpol} indicates
the critical role of the top electrode in the enhancement of the variation
in the ionic polarization among the stable configurations. To investigate
the effect of the electrode, we have computed the electron redistribution
for SrTiO$_{3}$/ZrO$_{2}$/Si stacks with 1.5 u.c.-thick STO, defined
as $\Delta n\left(x,y,z\right)=n_{\text{Au/STO/ZrO}_{2}\text{/Si}}\left(x,y,z\right)-n_{\text{STO/ZrO}_{2}\text{/Si}}\left(x,y,z\right)-n_{\text{Au}}\left(x,y,z\right)$.
Averaging o\textcolor{black}{ver the $1\times$ direction, we obtain
$\overline{\Delta n}\left(x,z\right)$, which is shown in \figref{SiSTO_ecd}(b)
and \figref{SiSTO_ecd}(d) for $S3$ and $S6$, respectively. In the
figure, it is observed that in both configurations, a charge transfer
from the oxide to the electrode occurs. This charge transfer is observed
to be slightly more pronounced in $S6$. To quantify the charge transfer,
we further average $\overline{\Delta n}\left(x,z\right)$ over the
$2\times$ direction to obtain $\overline{\Delta n}\left(z\right)$.
Computing the integral of $\overline{\Delta n}\left(z\right)$ up
to the electrode/oxide interface yields the total electron transfer
per $2\times1$ unit cell, which are 0.21, 0.23, 0.22, 0.22, 0.23
for $S1$ through $S5$, respectively, and 0.26 for $S6$. Plots analogous
to \figref{SiSTO_ecd} for the remaining configurations are presented
in the Supplementary Material.}

\begin{figure}
\begin{centering}
\includegraphics[width=0.9\columnwidth]{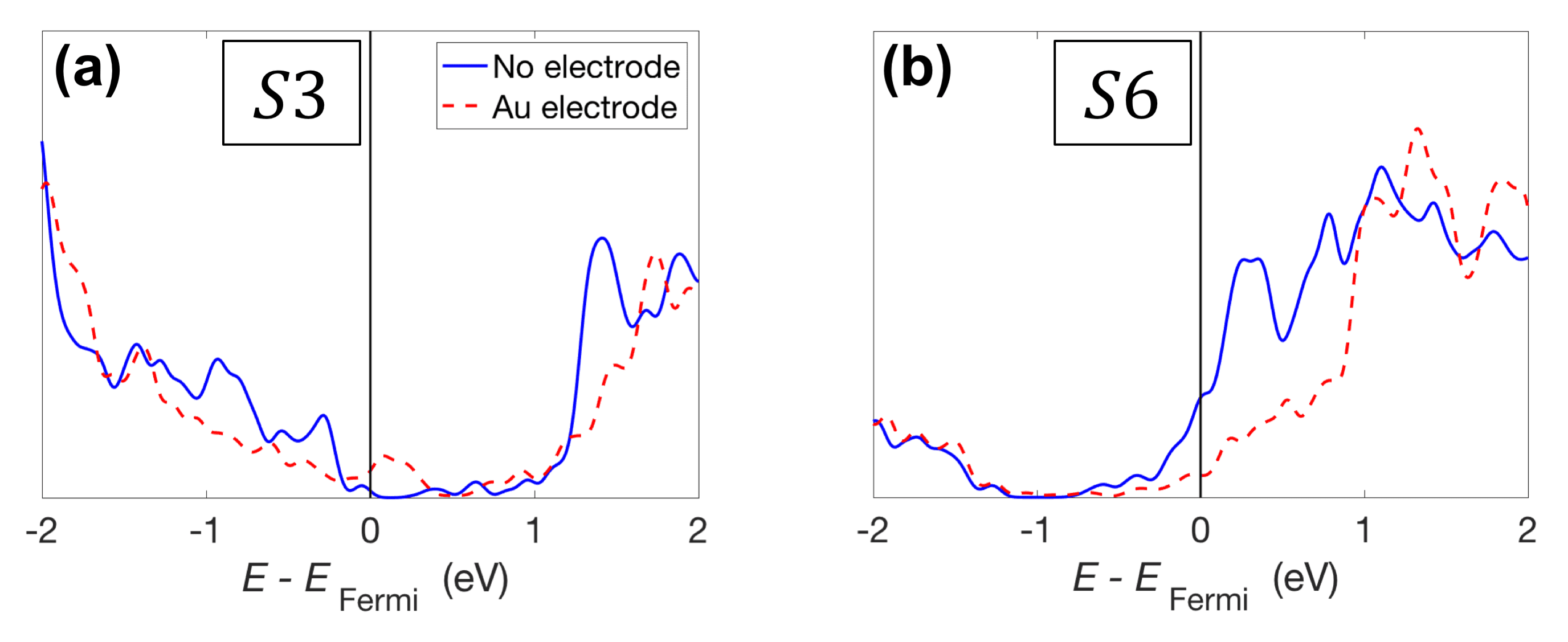}
\par\end{centering}
\caption{\label{fig:SiSTO_DOS}Densities of states (DOS) for (1.5 u.c. SrTiO$_{3}$)/ZrO$_{2}$/Si
heterostructures with and without the top Au electrode for (a) $S3$
and (b) $S6$ configurations. These DOS is approximated by summing
the projected DOS (PDOS) for all Si, O, Zr, Sr and Ti Löwdin orbitals
in the stack (excluding Au if present). The zero of the energy is
taken as the Fermi level for each case.}
\end{figure}

\textcolor{black}{To further illustrate the chemical difference between
the two types of interfaces ($S1$ through $S5$ versus $S6$) and
the effect of the electrode, we have computed densities of states
for (1.5 u.c. SrTiO$_{3}$)/ZrO$_{2}$/Si with and without the capping
Au electrode. Our results are presented in \figref{SiSTO_DOS}. In
order to exclude the electrode states and focus on the oxide/semiconductor
stack itself, we have summed the PDOS for all the Si, O, Zr, Sr and
Ti atoms. We see that prior to the addition of the electrode, $S3$
and $S6$ have similar DOS curves except for the position of the Fermi
level: $S3$ is a small-gap semiconductor whereas the Fermi level
of $S6$ is in the conduction band. When the electrode is added, $S3$
becomes lightly hole doped ($h_{1}$ loses electrons and becomes partially
filled when the electrode is added). For $S6$, the addition of the
electrode suppresses the DOS at and around the Fermi level. Therefore
the electron transfer mechanism in these two configurations are different:
hole doping of interfacial states in $S3$ and reduction in the electron
doping of the conduction band in $S6$. The plots corresponding to
the remaining configurations are presented in the Supplementary Material.
Because $S3$ and $S6$ are the two lowest-energy configurations (\figref{SiSTO_energy})
and have a large difference in ionic polarization (\figref{SiSTO_averpol}),
ferroelectric switching between them is in principle possible. According
to our analysis, this switching would be accompanied by a modulation
of the interfacial chemistry, between hole doping and electron doping,
analogous to the BaTiO$_{3}$/Ge interface \citep{dogan2017abinitio}}.

\section{Conclusion\label{sec:Conclusion5}}

We have conducted a density functional theory investigation of the
oxygen content for ZrO$_{x}$ monolayers o\textcolor{black}{n silicon
where $x=1.0,1.5,2.0,2.5,3.0$. We have found that the multiplicity
of low-energy structures obtained in ZrO$_{2}$ is preserved for both
under- and over-oxygenated monolayers. Our results indicate that ZrO$_{1.5}$
and ZrO$_{2.5}$ may also be used as a ferroelectric oxide with potentially
larger polarization switching. We have also presented our examination
of ZrO$_{2}$ as a buffer layer between Si and SrTiO$_{3}$. We have
found that SrTiO$_{3}$/ZrO$_{2}$/Si systems possess multiple atomic
configurations with different polarization profiles within a 1 eV
energy window. We have also shown that the relative energies of these
structures can be significantly changed with the help of a top electrode.
Using an electrode with the right work function, two low-energy structures
can be brought energetically close, which may allow for polarization
switching that does not involve high-energy metastable structures.}
This suggests that ZrO$_{2}$ may be used as an atomically-thin buffer
layer to induce switchable polarization in a thicker perovskite film
on silicon.\textcolor{black}{{} If this system is experimentally realized
and shown to have the desired transport properties, it will present
an attractive alternative in the field of non-volatile devices. The
main trends in our report may also guide future theoretical and experimental
research into monolayer oxides and their heterostructures.}

\section{Acknowledgements}

This work was supported primarily by the grant NSF MRSEC DMR-1119826.
We thank the Yale Center for Research Computing for guidance and use
of the research computing infrastructure, with special thanks to Stephen
Weston and Andrew Sherman. Additional computational support was provided
by NSF XSEDE resources via Grant TG-MCA08X007.

\bibliographystyle{apsrev}
\bibliography{Citations}

\begin{thebibliography}{29}
\expandafter\ifx\csname natexlab\endcsname\relax\def\natexlab#1{#1}\fi
\expandafter\ifx\csname bibnamefont\endcsname\relax
  \def\bibnamefont#1{#1}\fi
\expandafter\ifx\csname bibfnamefont\endcsname\relax
  \def\bibfnamefont#1{#1}\fi
\expandafter\ifx\csname citenamefont\endcsname\relax
  \def\citenamefont#1{#1}\fi
\expandafter\ifx\csname url\endcsname\relax
  \def\url#1{\texttt{#1}}\fi
\expandafter\ifx\csname urlprefix\endcsname\relax\def\urlprefix{URL }\fi
\providecommand{\bibinfo}[2]{#2}
\providecommand{\eprint}[2][]{\url{#2}}

\bibitem[{\citenamefont{Dogan et~al.}(2018)\citenamefont{Dogan,
  Fernandez-Pe{\~n}a, Kornblum, Jia, Kumah, Reiner, Krivokapic, Kolpak,
  Ismail-Beigi, Ahn et~al.}}]{dogan2018singleatomic}
\bibinfo{author}{\bibfnamefont{M.}~\bibnamefont{Dogan}},
  \bibinfo{author}{\bibfnamefont{S.}~\bibnamefont{Fernandez-Pe{\~n}a}},
  \bibinfo{author}{\bibfnamefont{L.}~\bibnamefont{Kornblum}},
  \bibinfo{author}{\bibfnamefont{Y.}~\bibnamefont{Jia}},
  \bibinfo{author}{\bibfnamefont{D.~P.} \bibnamefont{Kumah}},
  \bibinfo{author}{\bibfnamefont{J.~W.} \bibnamefont{Reiner}},
  \bibinfo{author}{\bibfnamefont{Z.}~\bibnamefont{Krivokapic}},
  \bibinfo{author}{\bibfnamefont{A.~M.} \bibnamefont{Kolpak}},
  \bibinfo{author}{\bibfnamefont{S.}~\bibnamefont{Ismail-Beigi}},
  \bibinfo{author}{\bibfnamefont{C.~H.} \bibnamefont{Ahn}},
  \bibnamefont{et~al.}, \bibinfo{journal}{Nano Letters}
  \textbf{\bibinfo{volume}{18}}, \bibinfo{pages}{241} (\bibinfo{year}{2018}),
  ISSN \bibinfo{issn}{1530-6984},
  \urlprefix\url{http://dx.doi.org/10.1021/acs.nanolett.7b03988}.

\bibitem[{\citenamefont{Dogan and Ismail-Beigi}(2019)}]{dogan2019theoryof}
\bibinfo{author}{\bibfnamefont{M.}~\bibnamefont{Dogan}} \bibnamefont{and}
  \bibinfo{author}{\bibfnamefont{S.}~\bibnamefont{Ismail-Beigi}},
  \bibinfo{journal}{arXiv:1902.01022 [cond-mat]}  (\bibinfo{year}{2019}),
  \bibinfo{note}{arXiv: 1902.01022},
  \urlprefix\url{http://arxiv.org/abs/1902.01022}.

\bibitem[{\citenamefont{Hwang et~al.}(2012)\citenamefont{Hwang, Iwasa,
  Kawasaki, Keimer, Nagaosa, and Tokura}}]{hwang2012emergent}
\bibinfo{author}{\bibfnamefont{H.~Y.} \bibnamefont{Hwang}},
  \bibinfo{author}{\bibfnamefont{Y.}~\bibnamefont{Iwasa}},
  \bibinfo{author}{\bibfnamefont{M.}~\bibnamefont{Kawasaki}},
  \bibinfo{author}{\bibfnamefont{B.}~\bibnamefont{Keimer}},
  \bibinfo{author}{\bibfnamefont{N.}~\bibnamefont{Nagaosa}}, \bibnamefont{and}
  \bibinfo{author}{\bibfnamefont{Y.}~\bibnamefont{Tokura}},
  \bibinfo{journal}{Nature Materials} \textbf{\bibinfo{volume}{11}},
  \bibinfo{pages}{103} (\bibinfo{year}{2012}), ISSN \bibinfo{issn}{1476-1122},
  \urlprefix\url{http://www.nature.com/nmat/journal/v11/n2/abs/nmat3223.html}.

\bibitem[{\citenamefont{Mannhart and
  Schlom}(2010)}]{mannhart2010oxideinterfacestextemdashan}
\bibinfo{author}{\bibfnamefont{J.}~\bibnamefont{Mannhart}} \bibnamefont{and}
  \bibinfo{author}{\bibfnamefont{D.~G.} \bibnamefont{Schlom}},
  \bibinfo{journal}{Science} \textbf{\bibinfo{volume}{327}},
  \bibinfo{pages}{1607} (\bibinfo{year}{2010}), ISSN \bibinfo{issn}{0036-8075,
  1095-9203},
  \urlprefix\url{http://science.sciencemag.org/content/327/5973/1607}.

\bibitem[{\citenamefont{McKee et~al.}(2001)\citenamefont{McKee, Walker, and
  Chisholm}}]{mckee2001physical}
\bibinfo{author}{\bibfnamefont{R.~A.} \bibnamefont{McKee}},
  \bibinfo{author}{\bibfnamefont{F.~J.} \bibnamefont{Walker}},
  \bibnamefont{and} \bibinfo{author}{\bibfnamefont{M.~F.}
  \bibnamefont{Chisholm}}, \bibinfo{journal}{Science}
  \textbf{\bibinfo{volume}{293}}, \bibinfo{pages}{468} (\bibinfo{year}{2001}),
  ISSN \bibinfo{issn}{0036-8075, 1095-9203},
  \urlprefix\url{http://www.sciencemag.org/content/293/5529/468}.

\bibitem[{\citenamefont{Garrity et~al.}(2012)\citenamefont{Garrity, Kolpak, and
  Ismail-Beigi}}]{garrity2012growthand}
\bibinfo{author}{\bibfnamefont{K.~F.} \bibnamefont{Garrity}},
  \bibinfo{author}{\bibfnamefont{A.~M.} \bibnamefont{Kolpak}},
  \bibnamefont{and}
  \bibinfo{author}{\bibfnamefont{S.}~\bibnamefont{Ismail-Beigi}},
  \bibinfo{journal}{Journal of Materials Science}
  \textbf{\bibinfo{volume}{47}}, \bibinfo{pages}{7417} (\bibinfo{year}{2012}),
  ISSN \bibinfo{issn}{0022-2461, 1573-4803},
  \urlprefix\url{http://link.springer.com/article/10.1007/s10853-012-6425-z}.

\bibitem[{\citenamefont{Reiner et~al.}(2009)\citenamefont{Reiner, Walker, and
  Ahn}}]{reiner2009atomically}
\bibinfo{author}{\bibfnamefont{J.~W.} \bibnamefont{Reiner}},
  \bibinfo{author}{\bibfnamefont{F.~J.} \bibnamefont{Walker}},
  \bibnamefont{and} \bibinfo{author}{\bibfnamefont{C.~H.} \bibnamefont{Ahn}},
  \bibinfo{journal}{Science} \textbf{\bibinfo{volume}{323}},
  \bibinfo{pages}{1018} (\bibinfo{year}{2009}), ISSN \bibinfo{issn}{0036-8075,
  1095-9203}, \urlprefix\url{http://www.sciencemag.org/content/323/5917/1018}.

\bibitem[{\citenamefont{Reiner et~al.}(2010)\citenamefont{Reiner, Kolpak,
  Segal, Garrity, Ismail-Beigi, Ahn, and Walker}}]{reiner2010crystalline}
\bibinfo{author}{\bibfnamefont{J.~W.} \bibnamefont{Reiner}},
  \bibinfo{author}{\bibfnamefont{A.~M.} \bibnamefont{Kolpak}},
  \bibinfo{author}{\bibfnamefont{Y.}~\bibnamefont{Segal}},
  \bibinfo{author}{\bibfnamefont{K.~F.} \bibnamefont{Garrity}},
  \bibinfo{author}{\bibfnamefont{S.}~\bibnamefont{Ismail-Beigi}},
  \bibinfo{author}{\bibfnamefont{C.~H.} \bibnamefont{Ahn}}, \bibnamefont{and}
  \bibinfo{author}{\bibfnamefont{F.~J.} \bibnamefont{Walker}},
  \bibinfo{journal}{Advanced Materials} \textbf{\bibinfo{volume}{22}},
  \bibinfo{pages}{2919} (\bibinfo{year}{2010}), ISSN \bibinfo{issn}{1521-4095},
  \urlprefix\url{http://onlinelibrary.wiley.com/doi/10.1002/adma.200904306/abstract}.

\bibitem[{\citenamefont{Dogan and Ismail-Beigi}(2017)}]{dogan2017abinitio}
\bibinfo{author}{\bibfnamefont{M.}~\bibnamefont{Dogan}} \bibnamefont{and}
  \bibinfo{author}{\bibfnamefont{S.}~\bibnamefont{Ismail-Beigi}},
  \bibinfo{journal}{Physical Review B} \textbf{\bibinfo{volume}{96}},
  \bibinfo{pages}{075301} (\bibinfo{year}{2017}),
  \urlprefix\url{https://link.aps.org/doi/10.1103/PhysRevB.96.075301}.

\bibitem[{\citenamefont{Batra et~al.}(1973)\citenamefont{Batra, Wurfel, and
  Silverman}}]{batra1973phasetransition}
\bibinfo{author}{\bibfnamefont{I.~P.} \bibnamefont{Batra}},
  \bibinfo{author}{\bibfnamefont{P.}~\bibnamefont{Wurfel}}, \bibnamefont{and}
  \bibinfo{author}{\bibfnamefont{B.~D.} \bibnamefont{Silverman}},
  \bibinfo{journal}{Physical Review B} \textbf{\bibinfo{volume}{8}},
  \bibinfo{pages}{3257} (\bibinfo{year}{1973}),
  \urlprefix\url{http://link.aps.org/doi/10.1103/PhysRevB.8.3257}.

\bibitem[{\citenamefont{Dubourdieu et~al.}(2013)\citenamefont{Dubourdieu,
  Bruley, Arruda, Posadas, Jordan-Sweet, Frank, Cartier, Frank, Kalinin, Demkov
  et~al.}}]{dubourdieu2013switching}
\bibinfo{author}{\bibfnamefont{C.}~\bibnamefont{Dubourdieu}},
  \bibinfo{author}{\bibfnamefont{J.}~\bibnamefont{Bruley}},
  \bibinfo{author}{\bibfnamefont{T.~M.} \bibnamefont{Arruda}},
  \bibinfo{author}{\bibfnamefont{A.}~\bibnamefont{Posadas}},
  \bibinfo{author}{\bibfnamefont{J.}~\bibnamefont{Jordan-Sweet}},
  \bibinfo{author}{\bibfnamefont{M.~M.} \bibnamefont{Frank}},
  \bibinfo{author}{\bibfnamefont{E.}~\bibnamefont{Cartier}},
  \bibinfo{author}{\bibfnamefont{D.~J.} \bibnamefont{Frank}},
  \bibinfo{author}{\bibfnamefont{S.~V.} \bibnamefont{Kalinin}},
  \bibinfo{author}{\bibfnamefont{A.~A.} \bibnamefont{Demkov}},
  \bibnamefont{et~al.}, \bibinfo{journal}{Nature Nanotechnology}
  \textbf{\bibinfo{volume}{8}}, \bibinfo{pages}{748} (\bibinfo{year}{2013}),
  ISSN \bibinfo{issn}{1748-3387},
  \urlprefix\url{http://www.nature.com/nnano/journal/v8/n10/full/nnano.2013.192.html}.

\bibitem[{\citenamefont{Robertson}(2006)}]{robertson2006highdielectric}
\bibinfo{author}{\bibfnamefont{J.}~\bibnamefont{Robertson}},
  \bibinfo{journal}{Reports on Progress in Physics}
  \textbf{\bibinfo{volume}{69}}, \bibinfo{pages}{327} (\bibinfo{year}{2006}),
  ISSN \bibinfo{issn}{0034-4885},
  \urlprefix\url{http://iopscience.iop.org/0034-4885/69/2/R02}.

\bibitem[{\citenamefont{McDaniel et~al.}(2014)\citenamefont{McDaniel, Ngo,
  Posadas, Hu, Lu, Smith, Yu, Demkov, and Ekerdt}}]{mcdaniel2014achemical}
\bibinfo{author}{\bibfnamefont{M.~D.} \bibnamefont{McDaniel}},
  \bibinfo{author}{\bibfnamefont{T.~Q.} \bibnamefont{Ngo}},
  \bibinfo{author}{\bibfnamefont{A.}~\bibnamefont{Posadas}},
  \bibinfo{author}{\bibfnamefont{C.}~\bibnamefont{Hu}},
  \bibinfo{author}{\bibfnamefont{S.}~\bibnamefont{Lu}},
  \bibinfo{author}{\bibfnamefont{D.~J.} \bibnamefont{Smith}},
  \bibinfo{author}{\bibfnamefont{E.~T.} \bibnamefont{Yu}},
  \bibinfo{author}{\bibfnamefont{A.~A.} \bibnamefont{Demkov}},
  \bibnamefont{and} \bibinfo{author}{\bibfnamefont{J.~G.}
  \bibnamefont{Ekerdt}}, \bibinfo{journal}{Advanced Materials Interfaces}
  \textbf{\bibinfo{volume}{1}}, \bibinfo{pages}{n/a} (\bibinfo{year}{2014}),
  ISSN \bibinfo{issn}{2196-7350},
  \urlprefix\url{http://onlinelibrary.wiley.com/doi/10.1002/admi.201400081/abstract}.

\bibitem[{\citenamefont{McKee et~al.}(1998)\citenamefont{McKee, Walker, and
  Chisholm}}]{mckee1998crystalline}
\bibinfo{author}{\bibfnamefont{R.~A.} \bibnamefont{McKee}},
  \bibinfo{author}{\bibfnamefont{F.~J.} \bibnamefont{Walker}},
  \bibnamefont{and} \bibinfo{author}{\bibfnamefont{M.~F.}
  \bibnamefont{Chisholm}}, \bibinfo{journal}{Physical Review Letters}
  \textbf{\bibinfo{volume}{81}}, \bibinfo{pages}{3014} (\bibinfo{year}{1998}),
  \urlprefix\url{http://link.aps.org/doi/10.1103/PhysRevLett.81.3014}.

\bibitem[{\citenamefont{Kumah et~al.}(2016)\citenamefont{Kumah, Dogan, Ngai,
  Qiu, Zhang, Su, Specht, Ismail-Beigi, Ahn, and Walker}}]{kumah2016engineered}
\bibinfo{author}{\bibfnamefont{D.~P.} \bibnamefont{Kumah}},
  \bibinfo{author}{\bibfnamefont{M.}~\bibnamefont{Dogan}},
  \bibinfo{author}{\bibfnamefont{J.~H.} \bibnamefont{Ngai}},
  \bibinfo{author}{\bibfnamefont{D.}~\bibnamefont{Qiu}},
  \bibinfo{author}{\bibfnamefont{Z.}~\bibnamefont{Zhang}},
  \bibinfo{author}{\bibfnamefont{D.}~\bibnamefont{Su}},
  \bibinfo{author}{\bibfnamefont{E.~D.} \bibnamefont{Specht}},
  \bibinfo{author}{\bibfnamefont{S.}~\bibnamefont{Ismail-Beigi}},
  \bibinfo{author}{\bibfnamefont{C.~H.} \bibnamefont{Ahn}}, \bibnamefont{and}
  \bibinfo{author}{\bibfnamefont{F.~J.} \bibnamefont{Walker}},
  \bibinfo{journal}{Physical Review Letters} \textbf{\bibinfo{volume}{116}},
  \bibinfo{pages}{106101} (\bibinfo{year}{2016}),
  \urlprefix\url{http://link.aps.org/doi/10.1103/PhysRevLett.116.106101}.

\bibitem[{\citenamefont{Kolpak and
  Ismail-Beigi}(2011)}]{kolpak2011thermodynamic}
\bibinfo{author}{\bibfnamefont{A.~M.} \bibnamefont{Kolpak}} \bibnamefont{and}
  \bibinfo{author}{\bibfnamefont{S.}~\bibnamefont{Ismail-Beigi}},
  \bibinfo{journal}{Physical Review B} \textbf{\bibinfo{volume}{83}},
  \bibinfo{pages}{165318} (\bibinfo{year}{2011}),
  \urlprefix\url{http://link.aps.org/doi/10.1103/PhysRevB.83.165318}.

\bibitem[{\citenamefont{Xiao et~al.}(2019)\citenamefont{Xiao, Liu, Peng, Zheng,
  Feng, Zhang, Zhang, Hao, Liao, and g}}]{xiao2019performance}
\bibinfo{author}{\bibfnamefont{W.}~\bibnamefont{Xiao}},
  \bibinfo{author}{\bibfnamefont{C.}~\bibnamefont{Liu}},
  \bibinfo{author}{\bibfnamefont{Y.}~\bibnamefont{Peng}},
  \bibinfo{author}{\bibfnamefont{S.}~\bibnamefont{Zheng}},
  \bibinfo{author}{\bibfnamefont{Q.}~\bibnamefont{Feng}},
  \bibinfo{author}{\bibfnamefont{C.}~\bibnamefont{Zhang}},
  \bibinfo{author}{\bibfnamefont{J.}~\bibnamefont{Zhang}},
  \bibinfo{author}{\bibfnamefont{Y.}~\bibnamefont{Hao}},
  \bibinfo{author}{\bibfnamefont{M.}~\bibnamefont{Liao}}, \bibnamefont{and}
  \bibinfo{author}{\bibfnamefont{Y.~Z.} \bibnamefont{g}},
  \bibinfo{journal}{IEEE Electron Device Letters} pp. \bibinfo{pages}{1--1}
  (\bibinfo{year}{2019}), ISSN \bibinfo{issn}{0741-3106},
  \urlprefix\url{https://ieeexplore.ieee.org/document/8662670}.

\bibitem[{\citenamefont{Perdew et~al.}(1996)\citenamefont{Perdew, Burke, and
  Ernzerhof}}]{perdew1996generalized}
\bibinfo{author}{\bibfnamefont{J.~P.} \bibnamefont{Perdew}},
  \bibinfo{author}{\bibfnamefont{K.}~\bibnamefont{Burke}}, \bibnamefont{and}
  \bibinfo{author}{\bibfnamefont{M.}~\bibnamefont{Ernzerhof}},
  \bibinfo{journal}{Physical Review Letters} \textbf{\bibinfo{volume}{77}},
  \bibinfo{pages}{3865} (\bibinfo{year}{1996}),
  \urlprefix\url{http://link.aps.org/doi/10.1103/PhysRevLett.77.3865}.

\bibitem[{\citenamefont{Vanderbilt}(1990)}]{vanderbilt1990softselfconsistent}
\bibinfo{author}{\bibfnamefont{D.}~\bibnamefont{Vanderbilt}},
  \bibinfo{journal}{Physical Review B} \textbf{\bibinfo{volume}{41}},
  \bibinfo{pages}{7892} (\bibinfo{year}{1990}),
  \urlprefix\url{http://link.aps.org/doi/10.1103/PhysRevB.41.7892}.

\bibitem[{\citenamefont{Giannozzi et~al.}(2009)\citenamefont{Giannozzi, Baroni,
  Bonini, Calandra, Car, Cavazzoni, {Davide Ceresoli}, Chiarotti, Cococcioni,
  Dabo et~al.}}]{giannozzi2009quantum}
\bibinfo{author}{\bibfnamefont{P.}~\bibnamefont{Giannozzi}},
  \bibinfo{author}{\bibfnamefont{S.}~\bibnamefont{Baroni}},
  \bibinfo{author}{\bibfnamefont{N.}~\bibnamefont{Bonini}},
  \bibinfo{author}{\bibfnamefont{M.}~\bibnamefont{Calandra}},
  \bibinfo{author}{\bibfnamefont{R.}~\bibnamefont{Car}},
  \bibinfo{author}{\bibfnamefont{C.}~\bibnamefont{Cavazzoni}},
  \bibinfo{author}{\bibnamefont{{Davide Ceresoli}}},
  \bibinfo{author}{\bibfnamefont{G.~L.} \bibnamefont{Chiarotti}},
  \bibinfo{author}{\bibfnamefont{M.}~\bibnamefont{Cococcioni}},
  \bibinfo{author}{\bibfnamefont{I.}~\bibnamefont{Dabo}}, \bibnamefont{et~al.},
  \bibinfo{journal}{Journal of Physics: Condensed Matter}
  \textbf{\bibinfo{volume}{21}}, \bibinfo{pages}{395502}
  (\bibinfo{year}{2009}), ISSN \bibinfo{issn}{0953-8984},
  \urlprefix\url{http://stacks.iop.org/0953-8984/21/i=39/a=395502}.

\bibitem[{\citenamefont{Marzari et~al.}(1999)\citenamefont{Marzari, Vanderbilt,
  De~Vita, and Payne}}]{marzari1999thermal}
\bibinfo{author}{\bibfnamefont{N.}~\bibnamefont{Marzari}},
  \bibinfo{author}{\bibfnamefont{D.}~\bibnamefont{Vanderbilt}},
  \bibinfo{author}{\bibfnamefont{A.}~\bibnamefont{De~Vita}}, \bibnamefont{and}
  \bibinfo{author}{\bibfnamefont{M.~C.} \bibnamefont{Payne}},
  \bibinfo{journal}{Physical Review Letters} \textbf{\bibinfo{volume}{82}},
  \bibinfo{pages}{3296} (\bibinfo{year}{1999}),
  \urlprefix\url{http://link.aps.org/doi/10.1103/PhysRevLett.82.3296}.

\bibitem[{\citenamefont{Bengtsson}(1999)}]{bengtsson1999dipolecorrection}
\bibinfo{author}{\bibfnamefont{L.}~\bibnamefont{Bengtsson}},
  \bibinfo{journal}{Physical Review B} \textbf{\bibinfo{volume}{59}},
  \bibinfo{pages}{12301} (\bibinfo{year}{1999}),
  \urlprefix\url{http://link.aps.org/doi/10.1103/PhysRevB.59.12301}.

\bibitem[{\citenamefont{Miyamoto and Oshiyama}(1990)}]{miyamoto1990atomicand}
\bibinfo{author}{\bibfnamefont{Y.}~\bibnamefont{Miyamoto}} \bibnamefont{and}
  \bibinfo{author}{\bibfnamefont{A.}~\bibnamefont{Oshiyama}},
  \bibinfo{journal}{Physical Review B} \textbf{\bibinfo{volume}{41}},
  \bibinfo{pages}{12680} (\bibinfo{year}{1990}),
  \urlprefix\url{http://link.aps.org/doi/10.1103/PhysRevB.41.12680}.

\bibitem[{\citenamefont{Uchiyama and Tsukada}(1996)}]{uchiyama1996atomicand}
\bibinfo{author}{\bibfnamefont{T.}~\bibnamefont{Uchiyama}} \bibnamefont{and}
  \bibinfo{author}{\bibfnamefont{M.}~\bibnamefont{Tsukada}},
  \bibinfo{journal}{Surface Science}
  \textbf{\bibinfo{volume}{357{\textendash}358}}, \bibinfo{pages}{509}
  (\bibinfo{year}{1996}), ISSN \bibinfo{issn}{0039-6028},
  \urlprefix\url{http://www.sciencedirect.com/science/article/pii/0039602896800768}.

\bibitem[{\citenamefont{Kittel}(2004)}]{kittel2004introduction}
\bibinfo{author}{\bibfnamefont{C.}~\bibnamefont{Kittel}},
  \emph{\bibinfo{title}{Introduction to {Solid} {State} {Physics}}}
  (\bibinfo{publisher}{Wiley}, \bibinfo{year}{2004}), ISBN
  \bibinfo{isbn}{978-0-471-41526-8},
  \urlprefix\url{https://books.google.com/books?id=kym4QgAACAAJ}.

\bibitem[{\citenamefont{Heyd et~al.}(2005)\citenamefont{Heyd, Peralta,
  Scuseria, and Martin}}]{heyd2005energyband}
\bibinfo{author}{\bibfnamefont{J.}~\bibnamefont{Heyd}},
  \bibinfo{author}{\bibfnamefont{J.~E.} \bibnamefont{Peralta}},
  \bibinfo{author}{\bibfnamefont{G.~E.} \bibnamefont{Scuseria}},
  \bibnamefont{and} \bibinfo{author}{\bibfnamefont{R.~L.}
  \bibnamefont{Martin}}, \bibinfo{journal}{The Journal of Chemical Physics}
  \textbf{\bibinfo{volume}{123}}, \bibinfo{pages}{174101}
  (\bibinfo{year}{2005}), ISSN \bibinfo{issn}{0021-9606},
  \urlprefix\url{https://aip.scitation.org/doi/10.1063/1.2085170}.

\bibitem[{\citenamefont{Kolpak et~al.}(2010)\citenamefont{Kolpak, Walker,
  Reiner, Segal, Su, Sawicki, Broadbridge, Zhang, Zhu, Ahn
  et~al.}}]{kolpak2010interfaceinduced}
\bibinfo{author}{\bibfnamefont{A.~M.} \bibnamefont{Kolpak}},
  \bibinfo{author}{\bibfnamefont{F.~J.} \bibnamefont{Walker}},
  \bibinfo{author}{\bibfnamefont{J.~W.} \bibnamefont{Reiner}},
  \bibinfo{author}{\bibfnamefont{Y.}~\bibnamefont{Segal}},
  \bibinfo{author}{\bibfnamefont{D.}~\bibnamefont{Su}},
  \bibinfo{author}{\bibfnamefont{M.~S.} \bibnamefont{Sawicki}},
  \bibinfo{author}{\bibfnamefont{C.~C.} \bibnamefont{Broadbridge}},
  \bibinfo{author}{\bibfnamefont{Z.}~\bibnamefont{Zhang}},
  \bibinfo{author}{\bibfnamefont{Y.}~\bibnamefont{Zhu}},
  \bibinfo{author}{\bibfnamefont{C.~H.} \bibnamefont{Ahn}},
  \bibnamefont{et~al.}, \bibinfo{journal}{Physical Review Letters}
  \textbf{\bibinfo{volume}{105}}, \bibinfo{pages}{217601}
  (\bibinfo{year}{2010}),
  \urlprefix\url{http://link.aps.org/doi/10.1103/PhysRevLett.105.217601}.

\bibitem[{\citenamefont{Haeni et~al.}(2004)\citenamefont{Haeni, Irvin, Chang,
  Uecker, Reiche, Li, Choudhury, Tian, Hawley, Craigo
  et~al.}}]{haeni2004roomtemperature}
\bibinfo{author}{\bibfnamefont{J.~H.} \bibnamefont{Haeni}},
  \bibinfo{author}{\bibfnamefont{P.}~\bibnamefont{Irvin}},
  \bibinfo{author}{\bibfnamefont{W.}~\bibnamefont{Chang}},
  \bibinfo{author}{\bibfnamefont{R.}~\bibnamefont{Uecker}},
  \bibinfo{author}{\bibfnamefont{P.}~\bibnamefont{Reiche}},
  \bibinfo{author}{\bibfnamefont{Y.~L.} \bibnamefont{Li}},
  \bibinfo{author}{\bibfnamefont{S.}~\bibnamefont{Choudhury}},
  \bibinfo{author}{\bibfnamefont{W.}~\bibnamefont{Tian}},
  \bibinfo{author}{\bibfnamefont{M.~E.} \bibnamefont{Hawley}},
  \bibinfo{author}{\bibfnamefont{B.}~\bibnamefont{Craigo}},
  \bibnamefont{et~al.}, \bibinfo{journal}{Nature}
  \textbf{\bibinfo{volume}{430}}, \bibinfo{pages}{758} (\bibinfo{year}{2004}),
  ISSN \bibinfo{issn}{0028-0836},
  \urlprefix\url{http://www.nature.com/nature/journal/v430/n7001/full/nature02773.html}.

\bibitem[{\citenamefont{Kolpak and Ismail-Beigi}(2012)}]{kolpak2012interface}
\bibinfo{author}{\bibfnamefont{A.~M.} \bibnamefont{Kolpak}} \bibnamefont{and}
  \bibinfo{author}{\bibfnamefont{S.}~\bibnamefont{Ismail-Beigi}},
  \bibinfo{journal}{Physical Review B} \textbf{\bibinfo{volume}{85}},
  \bibinfo{pages}{195318} (\bibinfo{year}{2012}),
  \urlprefix\url{http://link.aps.org/doi/10.1103/PhysRevB.85.195318}.

\end{thebibliography}

\end{document}


\chapter*{Supplementary Material for ``Ferroelectric ZrO$_{2}$ monolayers
as buffer layers between SrTiO$_{3}$ and Si''}
\begin{center}
{\large{}Mehmet Dogan$^{1,2,3,4}$ and Sohrab Ismail-Beigi$^{1,2,5,6}$}\\
\par\end{center}

\begin{center}
$^{1}$Center for Research on Interface Structures and Phenomena,
Yale University, New Haven, Connecticut 06520, USA
\par\end{center}

\begin{center}
$^{2}$Department of Physics, Yale University, New Haven, Connecticut
06520, USA
\par\end{center}

\begin{center}
$^{3}$Department of Physics, University of California, Berkeley,
California 94720, USA
\par\end{center}

\begin{center}
$^{4}$Materials Science Division, Lawrence Berkeley National Laboratory,
Berkeley, California 94720, USA
\par\end{center}

\begin{center}
$^{5}$Department of Applied Physics, Yale University, New Haven,
Connecticut 06520, USA
\par\end{center}

\begin{center}
$^{6}$Department of Mechanical Engineering and Materials Science,
Yale University, New Haven, Connecticut 06520, USA
\par\end{center}

\section*{Low-energy ZrO$_{x}$ configurations}

\begin{figure}[H]
\begin{centering}
\includegraphics[width=1\columnwidth]{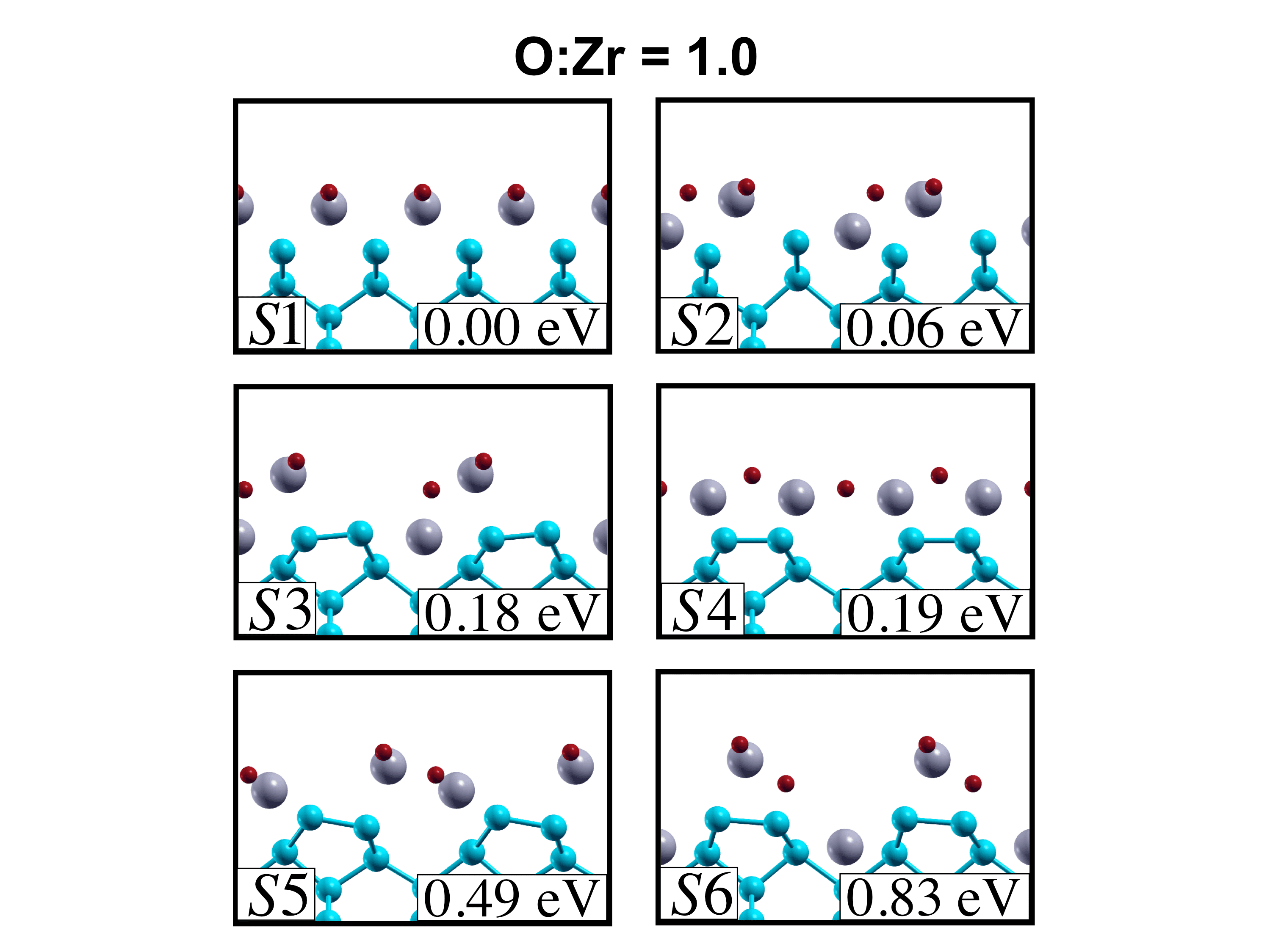}
\par\end{centering}
\caption{\label{fig:SiZrOx_10}Low-energy relaxed configurations of monolayers
with O:Zr = 1.0. Energies are per $2\times1$ in-plane cell.}
\end{figure}

\begin{figure}[H]
\begin{centering}
\includegraphics[width=1\columnwidth]{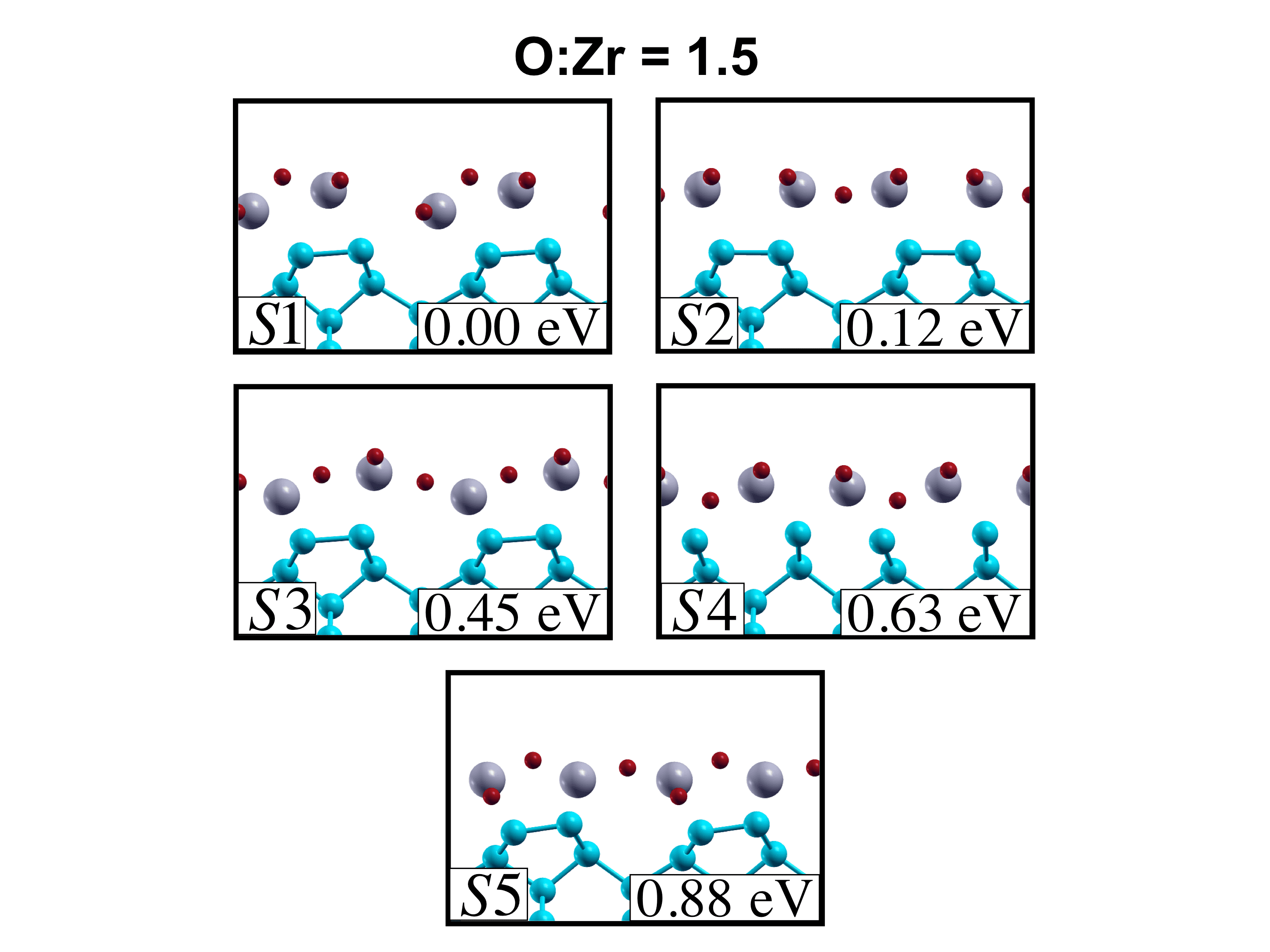}
\par\end{centering}
\caption{\label{fig:SiZrOx_15}Low-energy relaxed configurations of monolayers
with O:Zr = 1.5. Energies are per $2\times1$ in-plane cell.}
\end{figure}

\begin{figure}[H]
\begin{centering}
\includegraphics[width=1\columnwidth]{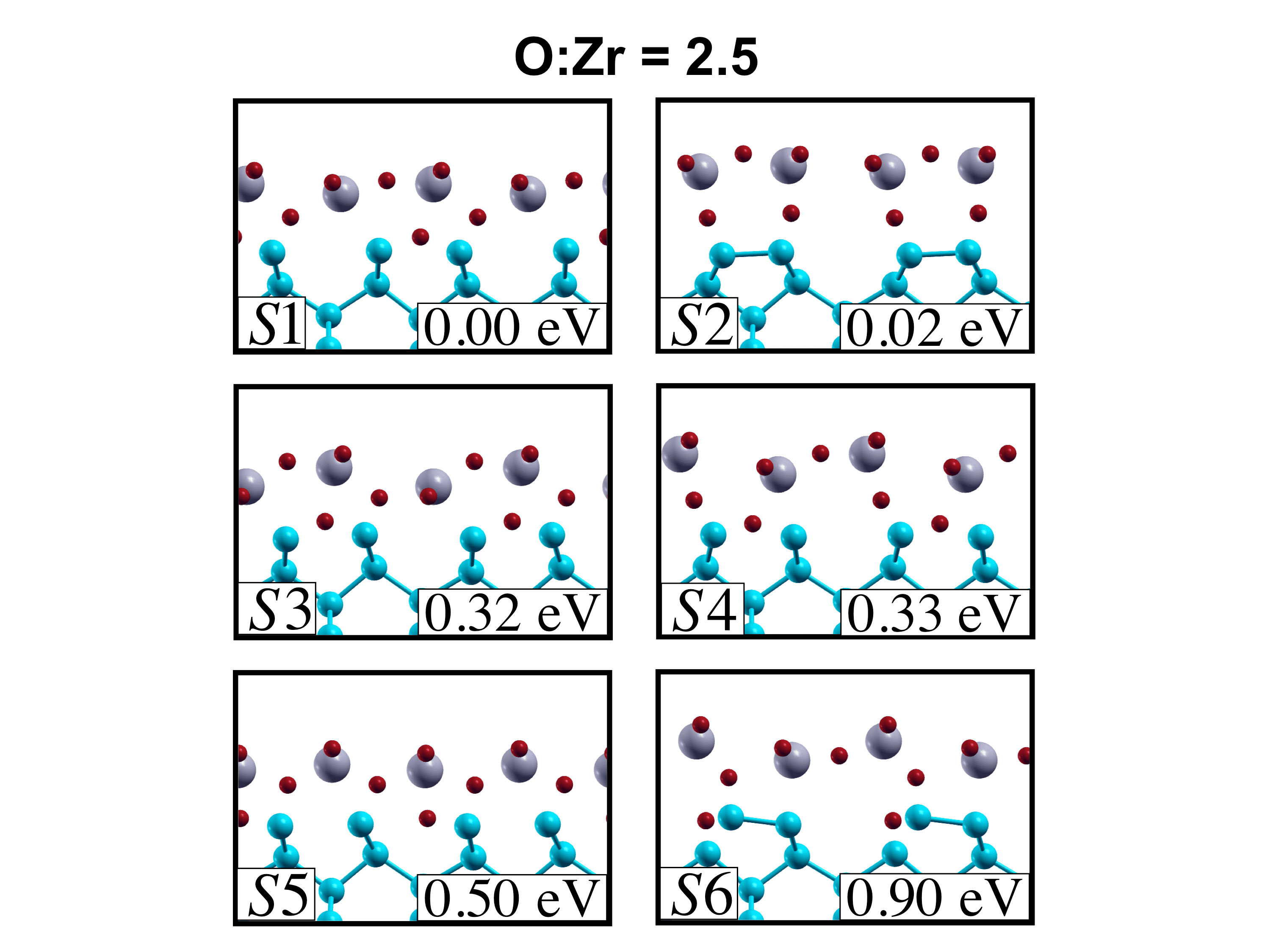}
\par\end{centering}
\caption{\label{fig:SiZrOx_25}Low-energy relaxed configurations of monolayers
with O:Zr = 2.5. Energies are per $2\times1$ in-plane cell.}
\end{figure}

\begin{figure}[H]
\begin{centering}
\includegraphics[width=1\columnwidth]{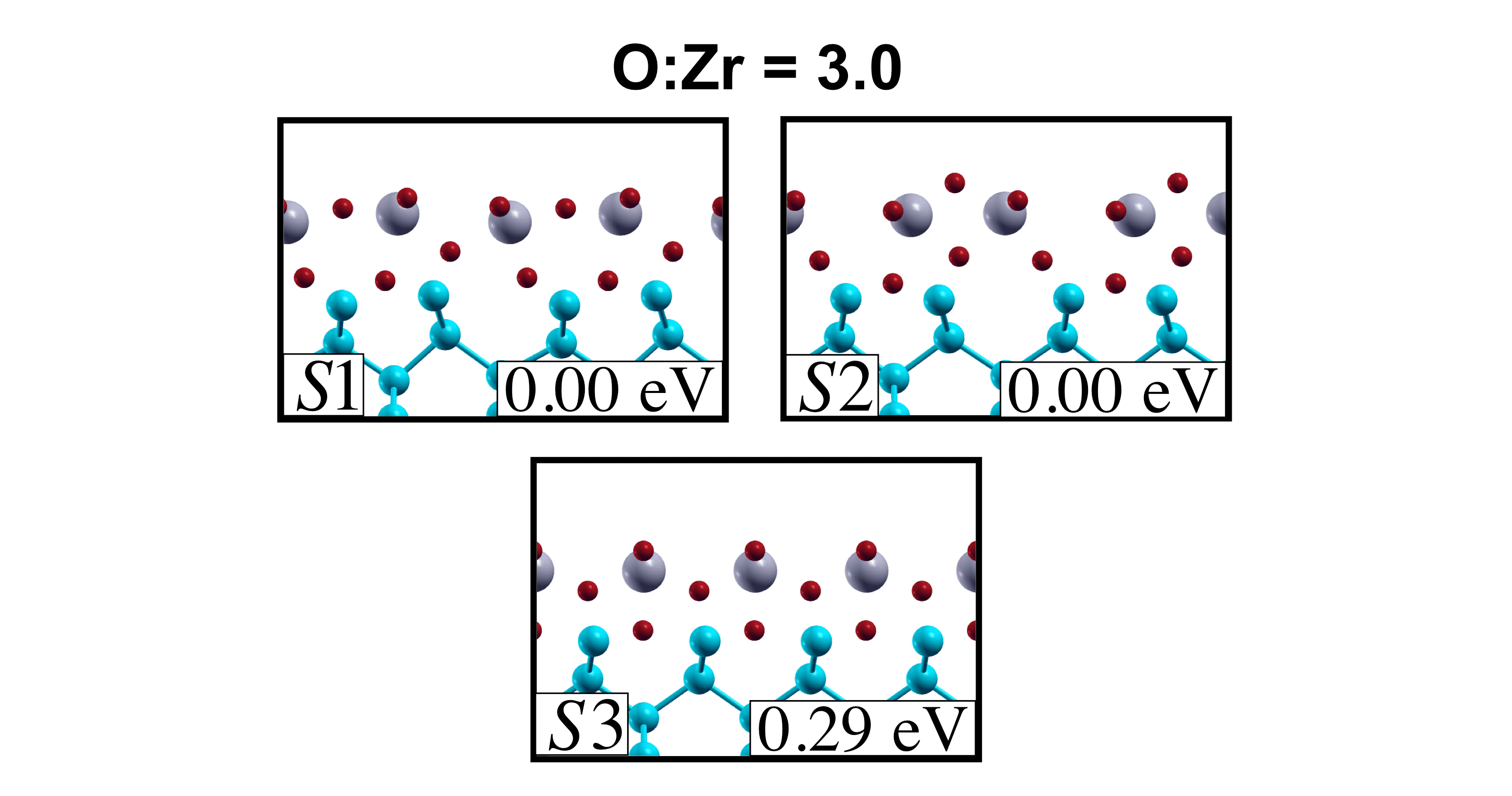}
\par\end{centering}
\caption{\label{fig:SiZrOx_30}Low-energy relaxed configurations of monolayers
with O:Zr = 3.0. Energies are per $2\times1$ in-plane cell.}
\end{figure}

\section*{Electronic structure of SrTiO$_{3}$/ZrO$_{2}$/Si stacks}

\begin{figure}[H]
\begin{centering}
\includegraphics[width=1\columnwidth]{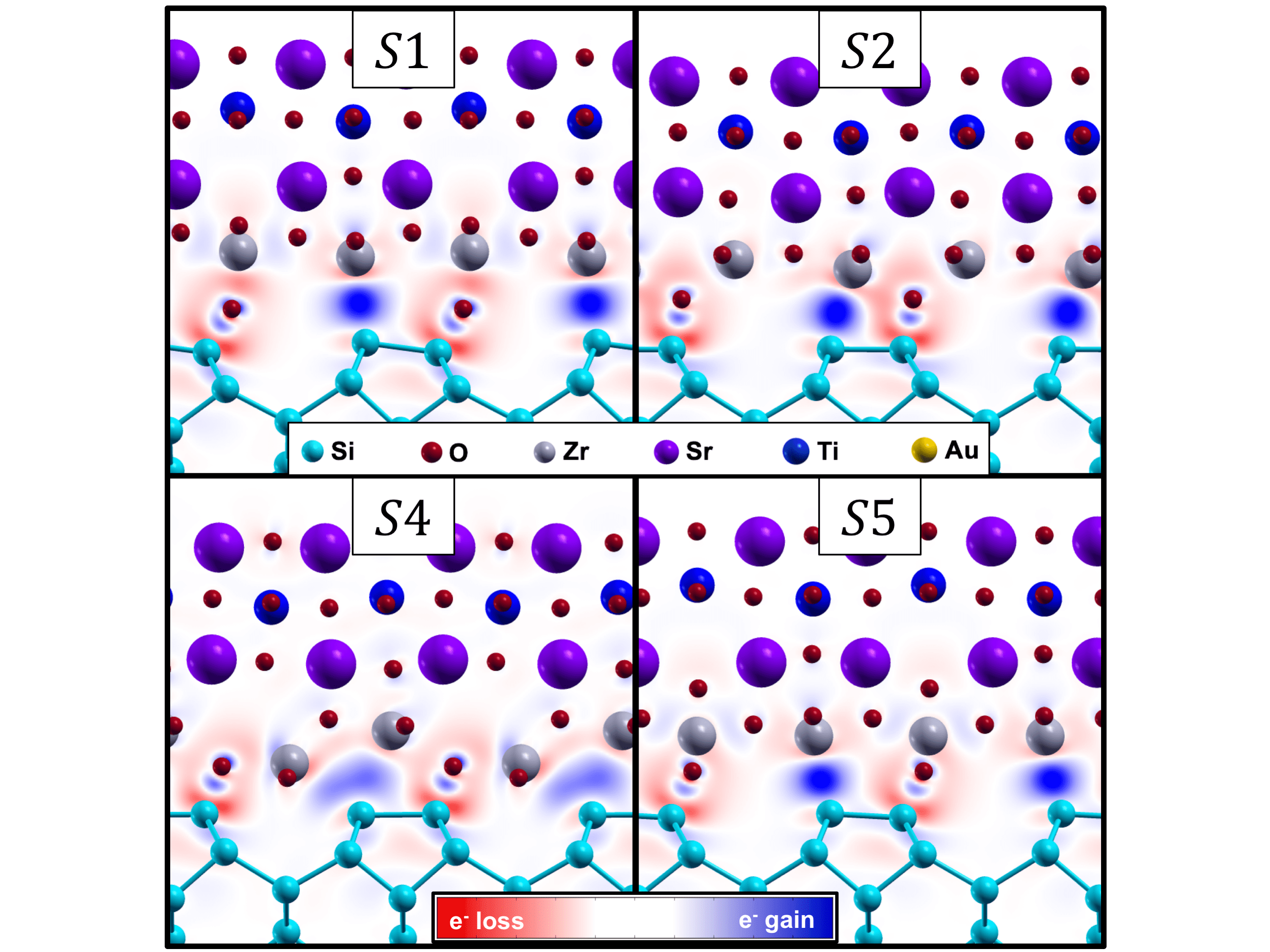}
\par\end{centering}
\caption{\label{fig:SiSTO_Sup_ECD1}Electron density redistribution for (1.5
u.c. SrTiO$_{3}$)/ZrO$_{2}$/Si heterostructures upon the formation
of the ZrO$_{2}$/Si interface for $S1$, $S2$, $S4$ and $S5$ configurations.}
\end{figure}

\begin{figure}[H]
\begin{centering}
\includegraphics[width=1\columnwidth]{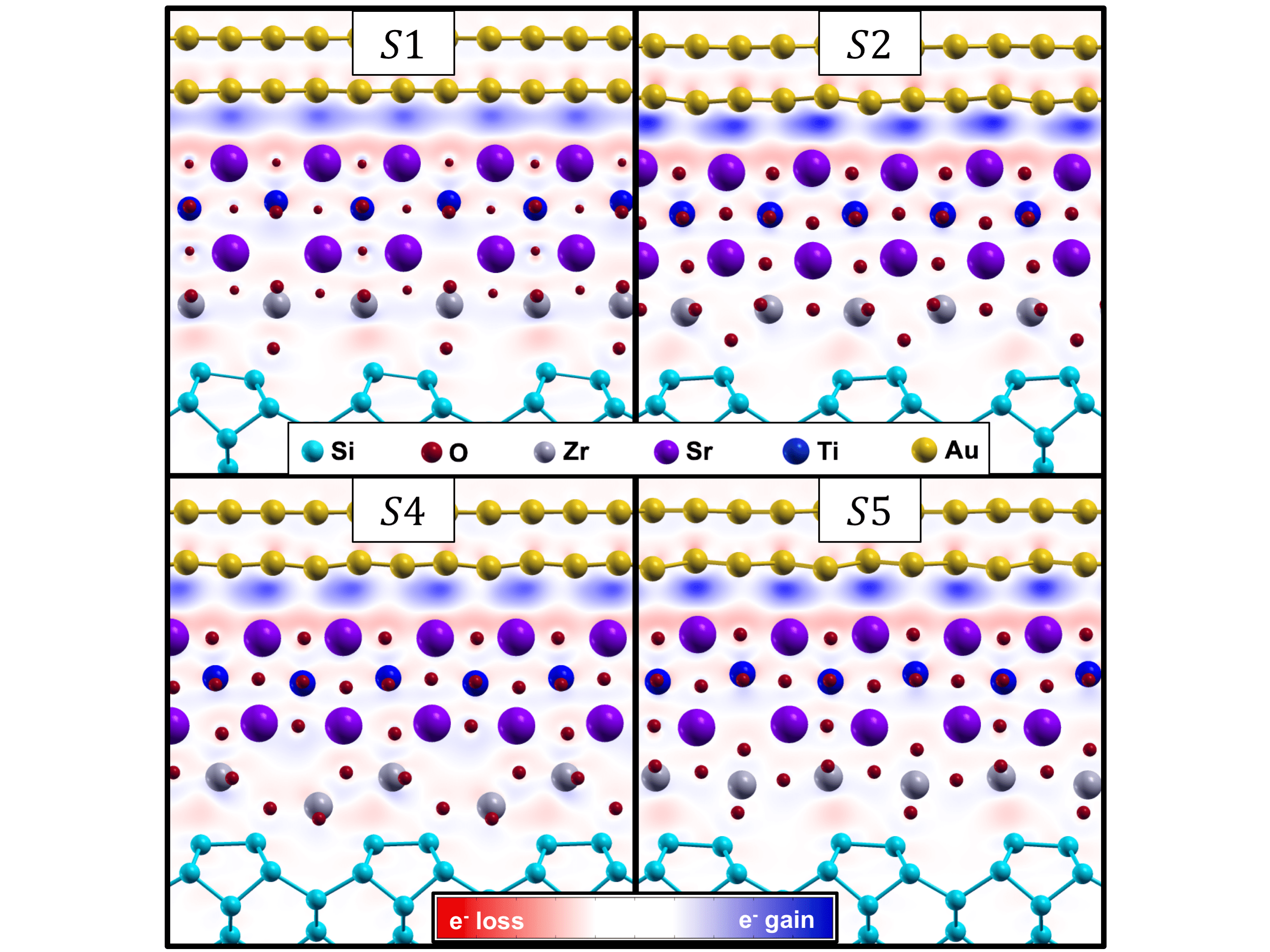}
\par\end{centering}
\caption{\label{fig:SiSTO_Sup_ECD2}Electron density redistribution for (1.5
u.c. SrTiO$_{3}$)/ZrO$_{2}$/Si heterostructures upon the addition
of the top Au electrode for $S1$, $S2$, $S4$ and $S5$ configurations.}
\end{figure}

\begin{figure}[H]
\begin{centering}
\includegraphics[width=1\columnwidth]{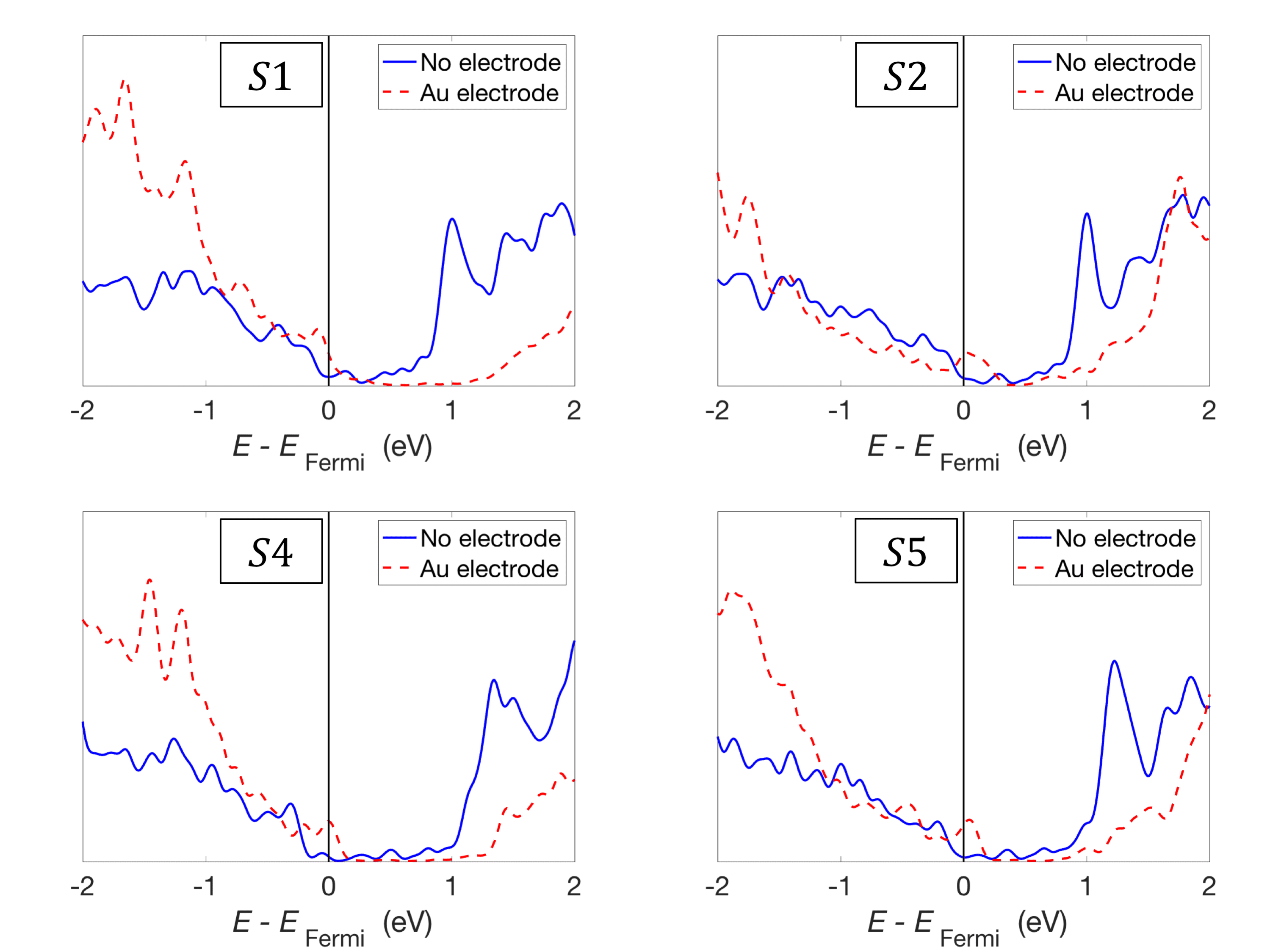}
\par\end{centering}
\caption{\label{fig:SiSTO_Sup_DOS}Densities of states (DOS) for (1.5 u.c.
SrTiO$_{3}$)/ZrO$_{2}$/Si heterostructures with and without the
top Au electrode for $S1$, $S2$, $S4$ and $S5$ configurations.
The total DOS is approximated by summing the PDOS for all the Si,
O, Zr, Sr and Ti atoms in the stack (thereby excluding the Au states).
The zero of the energy is taken as the Fermi level for each case.}
\end{figure}